%% file: main.tex
\documentclass[10pt]{article}

\include{header}

\begin{document}

\maketitle

\begin{abstract}
Finite element schemes based on discontinuous Galerkin methods possess features amenable to massively parallel computing accelerated with general purpose graphics processing units (GPUs).
However, the computational performance of such schemes strongly depends on their implementation.
In the past, several implementation strategies have been proposed.
They are based \modif{exclusively on specialized compute kernels} tuned for each operation, or they can leverage BLAS libraries that provide optimized routines for basic linear algebra operations.
In this paper, we present and analyze up-to-date performance results for different implementations, tested in a unified framework on a single NVIDIA GTX980 GPU.
We show that specialized kernels written with a one-node-per-thread strategy are competitive for polynomial bases up to the fifth and seventh degrees for acoustic and elastic models, respectively.
For higher degrees, a strategy that makes use of the NVIDIA cuBLAS library provides better results, able to reach a net arithmetic throughput 35.7\% of the theoretical peak value.
\end{abstract}




\section{Introduction}
\input{intro}


\section{Discontinuous Galerkin schemes}
\label{sec:schemod}

We consider acoustic and isotropic elastic wave models discretized with a nodal discontinuous Galerkin method in space and on a third-order Adam-Bashforth method in time.
We assume the physical coefficients to be constant over each mesh cell and discontinuous at interfaces.
The variational forms of the models and the schemes are presented in sections \ref{sec:models} and \ref{sec:schemes}, respectively.

\input{DGmodels}
\input{DGschemes}


\section{Implementations for GPU computing}
\label{sec:implements}

We consider three different GPU-based implementations of DG schemes.
Each has been programmed in a \CC\ code using CUDA 7.5 through the abstract framework OCCA \cite{Medina2014}, and tested with a single NVIDIA GTX980.
In section \ref{sec:imp:gen}, we summarize key characteristics of NVIDIA's GPU architecture and CUDA programming.
Specifications corresponding to the NVIDIA GTX980 are mentioned in brackets \textit{[...]}.
We present two implementations based \modif{exclusively on specialized  kernels} in section \ref{sec:imp:kernels}.
In section \ref{sec:imp:CUBLAS}, we propose an alternative strategy that makes use of the basic linear algebra subroutine SGEMM, from NVIDIA's cuBLAS library, optimized for dense algebraic operations on GPU.

\input{implementGeneralities}

\input{implementKernels}

\input{implementCUBLAS}


\section{Performance results}
\label{sec:perf}

The different implementations are tested and profiled in a basic setting: the time stepping of an initial solution on a given mesh.
We consider polynomial degrees ranging from $1$ to $8$ with, for each degree, a mesh corresponding to approx.\ 2 millions nodes.
\modif{All the results therefore correspond to the same number of discrete unknowns (approx.\ 8 and 18 millions for the acoustic and elastic models, respectively).
The computations are performed in single precision.}

In section \ref{sec:perf:gemm:padd}, we discuss the memory storage of the SGEMM routine for the DG operations requiered by the third implementation.
We then profile the SGEMM routine, report the arithmetic throughput and the memory bandwidth, and combine them in the roofline model (section \ref{sec:perf:gemm:prof}).
In sections \ref{sec:perf:DG:prof}, we profile the DG kernels of the three implementations for the acoustic case, and we compare the global performances of the implementations for both acoustic and elastic cases.

\input{perfGemmPadd}

\input{perfGemmProfile}

\input{perfDgProfile}


\section{Conclusion}
\input{conclusion}

\section*{Acknowledgements}

This work has been partly supported by NSF (award number DMS-1216674), Shell Global Solutions International B.V. and Shell Oil Company.
Axel Modave is partially supported by an excellence grant from Wallonie-Bruxelles International (WBI).
He is an Honorary Fellow of the Belgian American Educational Foundation (BAEF) and a Postdoctoral Researcher on leave with the F.R.S-FNRS.
The authors acknowledge Ty Mckercher and NVIDIA Corporation for providing GPU equipment.
The authors gratefully thank Jesse Chan for his helpful comments on the manuscript.


\bibliographystyle{abbrvnat}
\bibliography{myrefs}

\end{document}

%% file: header.tex
\usepackage{fullpage}
\usepackage[font=small,labelfont=small]{caption}

\newcommand\CC{C\nolinebreak[4]\hspace{-.05em}\raisebox{.4ex}{\relsize{-3}{\textbf{++}}}}

\usepackage{amsmath,amssymb,amsfonts}
\usepackage{stmaryrd}
\usepackage{xfrac}
\usepackage{eqnarray}
\usepackage{multirow}
\usepackage{colortbl}
\usepackage[table]{xcolor}

\usepackage{authblk}
\usepackage[numbers,sort&compress]{natbib}
\usepackage{subcaption}
\usepackage{hyperref}

\usepackage[ruled,vlined]{algorithm2e}

\usepackage{color}
\definecolor{gray}{gray}{0.5}
\definecolor{lightgray}{gray}{0.75}
\definecolor{lightlightgray}{gray}{0.9}

\newcommand{\diff}[2]{\frac{\partial #1}{\partial #2}}
\newcommand{\ddiff}[2]{\frac{d #1}{d #2}}

\newcommand{\D}{\mathsf{D}}

\newcommand{\fields}{\text{fields}}
\newcommand{\traces}{\text{traces}}

\newcommand{\sym}[1]{\mathrm{sym}\left({#1}\right)}

\newcommand{\eg}{\textit{e.g.} }

\newcommand{\etal}{\textit{et al.} }

\newcommand{\mean}[1]{\left\{{#1}\right\}}
\newcommand{\jump}[1]{\left\llbracket{#1}\right\rrbracket}

\newcommand{\mat}[1]{ \boldsymbol{\bf #1 } }
\renewcommand{\vec}[1]{ \boldsymbol{\bf #1 } }

\newcommand{\modif}[1]{{\color{black}#1}}

\title{GPU performance analysis of a nodal discontinuous Galerkin method for acoustic and elastic models}

\author[1]{A. Modave\footnote{Corresponding author: modave@vt.edu}}
\author[2]{A. St-Cyr}
\author[1]{T. Warburton}
\affil[1]{Virginia Polytechnic Institute and State University, Blacksburg, Virginia, USA}
\affil[2]{Shell Global Solutions International B.V., Rijswijk, The Netherlands}

\date{}


%% file: intro.tex

High-performance compute resources are used intensively in modern computational seismology for applications including earthquake simulation and seismic imaging.
State-of-the-art compute clusters consist of massively parallel many-core central processing units (CPUs) with graphics processing units (GPUs) or \modif{coprocessors} to provide a performance boost.
Computing on GPUs has been made accessible to the scientific community thanks to programming frameworks that expand languages already widely used (\eg C, \CC\ or Fortran).
However, to fully leverage the capabilities of GPUs, applications must be developed taking into account specifics of each architecture.
It is therefore critical to develop computational algorithms that can be efficiently implemented on these architectures.


In computational seismology, several implementations have been proposed for numerical methods suitable for GPUs \modif{coprocessors}. Examples include tuned implementations for finite-difference schemes \citep{Michea2010,Weiss2013,Abdelkhalek2012,rubio2014finite} or finite-element schemes \citep{heinecke2014petascale,komatitsch2009porting,Komatitsch2010,Mu2013b,Modave2015a}.
Nowadays, the finite-difference schemes are the most widely used, and a large literature is available (see \eg Virieux \etal \cite{Virieux2011} for a review).
However, the stencil based reconstruction used in finite difference methods is not ideally suited to resolving wave propagation in realistic physical heterogeneous media typically suffering from loss of accuracy \cite{symes2009interface}.
By contrast, finite-element methods based on unstructured meshes are better suited to handle such material interfaces.
Several variants have been investigated, such as spectral finite-element methods \citep{Komatitsch1998, Komatitsch1999}, continuous mass-lumped finite-element methods \citep{Chin1999, Cohen2001} or discontinuous Galerkin (DG) methods \citep{Dumbser2006, deLaPuente2007, Collis2010, Etienne2010, Krebs2014, Mercerat2015}.
In contrast to the standard schemes, the mentioned methods do not require the solution of large sparse linear systems of equations, and it is possible to use explicit time-stepping schemes.
The DG approach, in particular, provides a framework that is both flexible for multi-scale modeling and appears to be well suited for GPU accelerated computing.
First, it can easily handle local time-stepping strategies \citep{warburton2008accelerating,Dumbser2009,Baldassari2011,Minisini2013}, and hybrid discretizations can be deployed by mixing different discretization orders and several kinds of elements \citep{Kirby2000,Dumbser2009,Etienne2010,Chan2015}.
Then, the weak element-to-element coupling and the dense algebraic operations required per element are suitable for parallel multi-threading computations with GPUs \citep{Klockner2009}.


Discontinuous Galerkin schemes can be implemented in different ways for GPU computing. Kl\"ockner \etal \cite{Klockner2009}  proposed an implementation for first-order wave systems discretized using a nodal DG scheme.
This implementation has been successfully adapted to several physical contexts \cite{Godel2010,Gandham2015,Modave2015a}, and we have recently presented an application for reverse-time migration with multi-rate time-stepping and GPU clusters \cite{Modave2015}.
Alternatively, Fuhry \etal \cite{Fuhry2014} have proposed an implementation for two-dimensional problems with a modal DG scheme.
All these implementations partition the computational work into tailored GPU kernels, which are optimized separately in order to improve performance.
A key difference between the approaches of Kl\"ockner \etal \cite{Klockner2009} and Fuhry \etal \cite{Fuhry2014} is the programming strategy of kernels: each thread performs computations corresponding to one node and to one element, respectively.
On the other hand, Witherden \etal \cite{witherden2014pyfr,witherden2015heterogeneous} recently developed an implementation based on a nodal DG method for applications of fluid dynamics.
This implementation, named PyFR, makes use of external BLAS libraries that provide linear algebra routines optimized for hardware devices.
Such a strategy has been successfully used for CPU computing \cite{chevaugeon2005efficient,Hesthaven2002,marchandise2006quadrature,vos2010h}.


In this paper, we investigate and evaluate three strategies to implement time-domain DG schemes for GPU computing.
All these strategies have been compared for three-dimensional acoustic and elastic cases with a unique computational framework.
They were programmed in \CC\ using CUDA 7.5 through the abstract framework OCCA \cite{Medina2014}, and tested on a single Nvidia GTX980.
We have tested one-element-per-thread and one-node-per-thread strategies, as well as a strategy that makes use of an external BLAS library.
In this work, we have used the generalized single precision matrix-multiplication (SGEMM) routine of NVIDIA's cuBLAS library \cite{cublas}.
We show that, similarly to CPU implementations \cite{vos2010h}, the best GPU implementation depends on the polynomial degree.
The one-node-per-thread tailored kernels provided the best runtime for small degree, while the implementation with SGEMM is better for higher degrees.


This paper is organized as follows.
In section \ref{sec:schemod}, we present the mathematical models for acoustic and elastic wave propagation, and we describe the nodal discontinuous Galerkin method and the time-stepping scheme.
In section \ref{sec:implements}, key aspects of GPU hardware and GPU programming are summarized, and all the implementations are described.
Section \ref{sec:perf} is dedicated to performance results and comparisons.
We discuss the optimization of the SGEMM routine for the DG operations, and we analyse its performance using the roofline model.
The implementations are then compared, and all the kernels are systematically profiled.

%% file: DGmodels.tex
\subsection{Physical models and variational forms}
\label{sec:models}

Acoustic waves are governed by the pressure-velocity system, which reads
\begin{eqnarray*}
  \diff{p}{t} + \rho c^2 \: \nabla\cdot\vec{v} &=& 0, \\
  \rho \diff{\vec{v}}{t} + \nabla p &=& 0,
\end{eqnarray*}
with the pressure $p(\vec{x},t)$, the velocity $\vec{v}(\vec{x},t)$, the density $\rho(\vec{x})$ and the phase velocity $c(\vec{x})$.
\modif{For each mesh cell $\D_k$}, we consider the variational form
\begin{eqnarray*}
  \int_{\D_k} \diff{p}{t}\:\psi \:d\vec{x}
  &=&
    -\int_{\D_k} \rho c^2\:\nabla\cdot\vec{v} \: \psi \:d\vec{x}
    -\int_{\partial \D_k} p_p \: \psi \: d\vec{x}
  \\
  \int_{\D_k} \diff{\vec{v}}{t}\:\psi\:d\vec{x}
  &=&
    -\int_{\D_k} \frac{1}{\rho}\nabla p \: \psi \:d\vec{x}
    -\int_{\partial \D_k} p_{\vec{v}}\vec{n} \: \psi \:d\vec{x}
\end{eqnarray*}
where $\psi(\vec{x})$ is a test function, $p_p(\vec{x})$ and $p_{\vec{v}}(\vec{x})$ are penalty terms.
The penalty terms corresponding to upwind fluxes provided by a one-dimensional Riemann solver are given by \modif{(see \eg \cite{Hesthaven2007,Modave2015})}
\begin{eqnarray*}
  p_p &=& - (\rho c^2)^- k_c\alpha_c, \\
  p_{\vec{v}} &=& c^- k_c\alpha_c,
\end{eqnarray*}
with $\alpha_c = \jump{p} - (\rho c)^+ \jump{v_n}$, $k_c=1/\mean{\rho c}$ and the semi-jumps defined as $\jump{p}=(p^+-p^-)/2$ and $\jump{v_n}=(\vec{v}^+-\vec{v}^-)\cdot\vec{n}/2$.
The brackets $\mean{\cdot}$ denotes the mean value at the interface.

For isotropic media, elastic waves can be simulated with the velocity-stress system, which reads
\begin{eqnarray*}
  \diff{\mat{\sigma}}{t} &=&
    \lambda (\nabla\cdot\vec{v}) \mat{I} 
    + 2 \mu\:\sym{\nabla\vec{v}}, \\
  \rho \diff{\vec{v}}{t} &=& \nabla\cdot\mat{\sigma},
\end{eqnarray*}
with the stress tensor $\mat{\sigma}(\vec{x},t)$ and the Lam\'e parameters $\lambda(\vec{x})$ and $\mu(\vec{x})$.
This system supports pressure waves and shear waves, which the phase velocities are
\begin{eqnarray*}
  c_p &=& \sqrt{(\lambda+2\mu)/\rho}, \\
  c_s &=& \sqrt{\mu/\rho},
\end{eqnarray*}
respectively.
We consider the variational form
\begin{eqnarray*}
  \int_{\D_k} \diff{\mat{\sigma}}{t} \: \psi \:d\vec{x}
  &=&
    \int_{\D_k} \lambda (\nabla\cdot\vec{v}) \mat{I} + 2 \mu\:\sym{\nabla\vec{v}} \: \psi \:d\vec{x}
  + \int_{\partial \D_k} \mat{P}_{\mat{\sigma}} \: \psi \:d\vec{x}, \\
  \int_{\D_k} \diff{\vec{v}}{t} \psi \:d\vec{x}
  &=&
    \int_{\D_k} \frac{1}{\rho}(\nabla\cdot\mat{\sigma}) \: \psi \:d\vec{x}
  + \int_{\partial \D_k} \vec{p}_{\vec{v}} \: \psi \:d\vec{x}
\end{eqnarray*}
where the upwind fluxes provided by an exact Riemann solver are given by \cite{Wilcox2010}
\begin{eqnarray*}
  \mat{P}_{\mat{\sigma}}
    &=& k_p \alpha_p \lambda^- \mat{I}
     + 2 \left(k_p \alpha_p - k_s \alpha_s\right) \mu^- \mat{N}
     + 2 k_s \mu^-\:\sym{\vec{n}\otimes\vec{\alpha}_s}, \\
  \vec{p}_{\vec{v}}
    &=& \left(k_p \alpha_p c_p^-
           - k_s \alpha_s c_s^-\right) \vec{n}
     + k_s c_s^- \vec{\alpha}_s,
\end{eqnarray*}
with
\begin{align*}
  \alpha_p &= \jump{\sigma_{nn}} + (\rho c_p)^+ \jump{v_n}, &
  k_p &= 1/\mean{\rho c_p}, &
  \jump{\sigma_{nn}} &= \vec{n}\cdot(\mat{\sigma}^+-\mat{\sigma}^-)\cdot\vec{n}/2, \\
  \alpha_s &= \jump{\sigma_{nn}} + (\rho c_s)^+ \jump{v_n}, &
  k_s &= 1/\mean{\rho c_s}, &
  \jump{\mat{\sigma}\vec{n}} &= (\mat{\sigma}^+-\mat{\sigma}^-)\cdot\vec{n}/2, \\
  \vec{\alpha}_s &= \jump{\mat{\sigma}\vec{n}} + (\rho c_s)^+ \jump{\vec{v}}, &
  \mat{N} &= \vec{n}\otimes\vec{n}, &
  \jump{\vec{v}} &= (\vec{v}^+-\vec{v}^-)/2.
\end{align*}

%% file: DGschemes.tex
\subsection{Numerical schemes}
\label{sec:schemes}

The approximate fields are built on a spatial mesh of the computational domain $\Omega \subset \mathbb{R}^3$, made of $K$ non-overlapping tetrahedral cells, $\Omega = \bigcup_k \D_k$, where $\D_k$ is the $k^\text{th}$ cell.
For the nodal discontinuous Galerkin method, all the scalar fields and the Cartesian components of vector and tensor fields are approximated by piecewise polynomial functions.
The discrete unknowns correspond to the values of fields at nodes distributed over the surface and interior of an element \cite{Hesthaven2002,Hesthaven2007}.
Over each cell $\D_k$, the approximate fields can then be represented by
\begin{equation}
  \vec{q}_k(\vec{x},t) = \sum_{n=1}^{N_p} \vec{q}_{k,n}(t)\;\ell_{k,n}(\vec{x}),
  \quad\forall\vec{x}\in \D_k,
\end{equation}
where $N_p$ is the number of nodes per element, $\vec{q}_{k,n}(t)$ is the values of fields at node $n$ of element $k$, and $\ell_{k,n}(\vec{x})$ is the associated multivariate Lagrange polynomial function.
In this work, the position of nodes are chosen using the warp-and-blend technique \cite{Warburton2006}.
The number of nodes for each element is given by $N_p = (N+1)(N+2)(N+3)/6$, where $N$ is the maximal order of polynomial functions. 

The semi-discrete equations are obtained by substituting the approximate representation of fields into the weak form, and using the Lagrange polynomials as test functions.
For each element $k$, a system of equations can then be expressed as
\begin{equation}
  \ddiff{\vec{q}_k}{t} = \vec{r}_{k},
  \label{eqn:dgLocalScheme}
\end{equation}
where the right-hand side vector is defined as
\begin{equation}
  \vec{r}_{k}
    = \sum_{i=1}^{3} \sum_{j=1}^3
      g_{k,i,j}^{\text{vol}} \vec{D}_{j} \vec{f}_{j,k}
    + \sum_{f=1}^{N_\text{faces}}
      g_{k,f}^{\text{sur}} \mat{L}_{f} \vec{p}_{k,f},
  \label{eqn:dgLocalRhs}
\end{equation}
where \modif{$N_\text{faces}$ is the number of faces,} $\vec{f}_{j,k}$ correspond to the physical flux in the $x_j$-direction for all the nodes of element $k$, and the vector $\vec{p}_{k,f}$ contains the penalty terms for all the nodes belonging to face $f$.
The first term of the right-hand side vector in equation \eqref{eqn:dgLocalRhs} corresponds to the volume integrals of the variational form, while the second term corresponds to the surface integrals.
The matrices $\mat{D}_{j}$ and $\mat{L}_{f}$ are respectively the differentiation matrices and the lifting matrices of the reference element.
The geometric factors $g_{k,i,j}^{\text{vol}}$ and $g_{k,f}^{\text{sur}}$ depend on the shape of each element.
The definitions of these matrices and factors, and a complete derivation of the semi-discrete equations are given in \cite{Modave2015}.

The global semi-discrete scheme is simply built by combining the local systems of all elements.
We use the classical third-order Adams-Bashforth formula for time discretization.
At each iteration, the discrete unknowns of all the elements are updated according to
\begin{equation}
    \vec{q}_k^{m+1}
  = \vec{q}_k^{m}
  + {\color{black}\Delta t}
    \sum_{s=1}^3 a_s\:\vec{r}_k^{m-s+1},
  \label{eqn:updateScheme}
\end{equation}
with $a_1 = 23/12$, $a_2 = -16/12$ and $a_3 = 5/12$, and where $m$ is the time index.
This formula only depends on the right-hand side vector $\vec{r}_k$ at the three time levels $t^m,t^{m-1},t^{m-2}$.
For each step, this vector is thus computed with the values of unknowns local to the element, as well as those of the neighboring element at each interface.

%% file: implementGeneralities.tex
\subsection{GPU architecture and GPU programming}
\label{sec:imp:gen}

Heterogeneous computing systems consist of central processing units (CPUs) complemented with sidecar graphics processing units (GPUs), each with its own memory space.
The programming framework CUDA \cite{cuda,cheng2014professional} provides extensions to the C programming language to allocate/deallocate storage in the global memory of GPU, to manage memory transfers between the CPU memory and the GPU memory, and to invoke \textit{kernels} --- the codes that run on the GPU.
In our implementations, memory allocations and CPU-GPU data transfers are performed only at the beginning of the computation, and during the time stepping only when the solution must be exported for checkpointing or halo exchanges.
The main challenge consists in optimizing the execution of tasks to be performed by the GPU.

The execution is based on the \textit{single instruction multiple threads} (SIMT) model, where many \textit{threads} are executed concurrently by the GPU.
In the programming framework CUDA, threads are grouped into \textit{thread blocks} \text{[max. 1024 threads per thread block]}, which compose the \textit{grid}.
At the hardware level, each \textit{thread block} is assigned to one \textit{streaming multiprocessor} (SM) as the kernel executes {[16 SMs on chip with 128 cores per SM]}, and successive threads of each thread block are grouped into \textit{warps} \text{[32 threads per warp]}.
Ideally, all the threads of a warp must perform the same instructions to avoid \textit{warp divergence}.
Warp schedulers manage the execution of warps, eventually context switching between warps to avoid the SM from  stalling \text{[max. 64 active warps per SM]} due to high device memory load and store latencies.
The percentage of active warps to the maximum theoretical number of active warps is called \textit{occupancy} . 
It is generally advised that the number of threads per thread block should be chosen to be a multiple of 32 to maximize the occupancy \cite{cudaPractice}, though, strategies with low occupancy can also be efficient \cite{volkov2010better}.

The GPU has a deep non-uniform memory architecture with a hierarchy of memory spaces, each with different purposes and characteristics that indicate their best use cases.
\textit{Global memory} \text{[4GB]} is the largest on device memory space and is accessible by all the threads, in all warps, in all thread-blocks, in all kernels.
The bandwidth of this memory \text{[max.\ 224GB/s]} is generally a bottleneck.
Common strategies to maximize usage of global memory include hiding latency due to data transfer by overlapping memory load and store requests with computations, and making coalesced memory transfers by requesting contiguous blocks of data that are page aligned to maximize the utilization of the memory bus.
Similarly to CPUs, GPUs have also a system of lower-latency \textit{caches} (L2 cache \text{[2MB]} and L1 cache \text{[48KB per SM]}), which allow reuse of data previously fetched from the global memory.
These caches cannot be programmed at the level of CUDA, and their use is determined by the streaming multiprocessor cache management system.

\textit{Shared memory} (smem) \text{[96KB per SM, max. 48KB per thread block]} is a relatively low latency memory that can be accessed by threads of a thread block, providing a scratch pad for thread-block level data sharing and collaboration.
Shared memory can be considered as a programmable cache, where memory storage is allocated at the thread block execution, and released when it is finished.
Shared memory accesses are served by 32 independent memory managers. Shared memory is organized into banks that are interleaved in the logical indexing of the memory addresses. Efficient  data transactions are served by the shared memory managers if all threads in a warp that access shared memory do so by accessing the 32 banks.
Finally, data that are local to each thread are stored in 32-bit \textit{registers} \text{[64K per SM, max.\ 255 per thread]} that provide very low latency accesses.
When the number of registers is insufficient to contain all the register variables for all the threads in a thread-block the \textit{spilled register} data are stored the local memory space, that is mapped to global memory and cached in L1 and L2 increasing the latency.
The storage allocated for spilling registers is called \textit{local memory} (lmem).
NVIDIA's GPUs also have constant memory and texture memory, but they are not used in this work.


%% file: implementKernels.tex
\subsection{Implementations with specialized kernels}
\label{sec:imp:kernels}

The operations to perform in the DG implementations can be partitioned into several kernels, which provides flexibility to optimize each task considering its own characteristics.
For each time-step update, the right-hand side vector \eqref{eqn:dgLocalRhs} must be computed at the current time, and the values of fields must be updated with the update scheme \eqref{eqn:updateScheme}.
Following previous works \cite{Godel2010, Fuhry2014, Gandham2015, Modave2015}, we consider implementations with three main kernels: the \textit{volume kernel} and the \textit{surface kernel} compute respectively the volume terms and the surface terms of the right-hand side vector, while the \textit{update kernel} performs the time stepping.
\modif{A similar partition of operations into tailored kernels has also been successfully tested for computing with coprocessors \cite{heinecke2014petascale}.}
We have tested implementations based on two different strategies: in all the kernels, each thread deals with the computations corresponding to one element (\textit{one-element-per-thread} strategy) or one node (\textit{one-node-per-thread} strategy).
These implementations are based on the work of Fuhry \etal \cite{Fuhry2014} and Kl\"ockner \etal \cite{Klockner2009}.
In the following we describe key aspects of memory storage and kernel programming with these strategies.

For both considered implementations, all the required data are stored in global memory: the array $\mathtt{Q}$ with the values of fields at all the nodes, the array $\mathtt{Qf}$ with the values of traces ($p$ and $\vec{n}\cdot\vec{u}$ for the acoustic case, and $\vec{u}$ and $\mat{\sigma}\vec{n}$ for the elastic case) at all the face nodes, the arrays $\mathtt{R}_0$, $\mathtt{R}_{-1}$ and $\mathtt{R}_{-2}$ with the values of the right-hand-side vector corresponding to the current and two previous time steps, the arrays $\mathtt{D}_i$ and $\mathtt{L}$ with the differentiation and lift matrices, and specific arrays for the geometric factors and physical coefficients.
The granularity of storage depends on the implementation strategy.
With the {one-element-per-thread} strategy, successive threads have to fetch/store values corresponding to successive elements.
It then is advantageous to store all the arrays with the element index $k$ as the finest level of granularity, which enables coalescing data transfers with global memory \cite{Fuhry2014}.
By contrast, with the {one-node-per-thread} strategy, successive threads deal with successive nodes.
A granularity with the node index $n$ at the finest level then is the best choice \cite{Modave2015}.
Dimensions and granularity of the main arrays are given in Table \ref{tab:arrays}.

\begin{table}
\centering
\begin{tabular}{l|l|l|l} \hline
\multirow{2}{*}{Name} & \multirow{2}{*}{Symbol} & Granularity & Granularity \\
     &        & \textit{one-element-per-thread} & \textit{one-node-per-thread} \\ \hline
Values of fields at nodes & $\mathtt{Q}$ &
    $N_{\fields} \cdot N_{p} \cdot K$ &
    $K \cdot N_{\fields} \cdot N_{p}$ \\
Values of traces at face nodes & $\mathtt{Qf}$ & 
    $N_{\traces} \cdot N_\text{faces} \cdot N_{fp} \cdot K$ & 
    $K \cdot N_{\traces} \cdot N_\text{faces} \cdot N_{fp}$ \\
Right-hand side terms & $\mathtt{R}_{\{0,-1,-2\}}$ &
    $N_{\fields} \cdot N_{p} \cdot K$ &
    $K \cdot N_{\fields} \cdot N_{p}$ \\
Differentiation matrices & $\mathtt{D}_i$ &
    $N_p^2$ &
    $N_p^2$ \\
Merged lift matrix & $\mathtt{L}$ &
    $N_\text{faces} \cdot N_{fp} \cdot N_p$ &
    $N_\text{faces} \cdot N_{fp} \cdot N_p$ \\ \hline
\end{tabular}
\caption{List of main arrays stored in the global memory of the device.
Dimensions are given from the coarsest to the finest granularity of storage.
$N_p$, $N_{fp}$, $N_\text{faces}$ and $K$ are respectively the number of nodes per element, the number of nodes per face, the number of faces per element, and the number of elements in the mesh.
$N_{\fields}$ and $N_{\traces}$ are the number of fields and traces, respectively.}
\label{tab:arrays}
\end{table}

We describe the tasks that must be performed by each kernel.
For the acoustic case, the right-hand side vector \eqref{eqn:dgLocalRhs} can be rewritten as
\begin{eqnarray}
  \vec{r}_k^p
    &=& -\rho_k c_k^2 \sum_{j=1}^{3} \vec{D}_j \left[\sum_{i=1}^{3} g_{k,i,j}^\text{vol} \vec{q}_{k}^{v_i}\right]
    - \mat{L}\vec{p}_{k}^p, \label{eqn:dgRhsAcou1} \\
  \vec{r}_k^{v_i}
    &=& - \frac{1}{\rho_k} \sum_{j=1}^3 g_{k,i,j}^\text{vol} \Big[\vec{D}_j \vec{q}_k^p\Big]
    - \mat{L}\vec{p}_{k}^{v_i}, \quad \text{for } i=1,2,3, \label{eqn:dgRhsAcou2}
\end{eqnarray}
where the lift matrices have been merged into $\mat{L}$, and the penalty vectors $\vec{p}_{k}^{\{p,v_i\}}$ contain the penalty terms multiplied with the geometric factors for all the face nodes of the element.
The superscripts $\:^p$ and $\:^{v_i}$ denote vectors with values corresponding to the pressure $p$ and the Cartesian component of velocity $v_i$.
Using similar notations, the right-hand side vector of the elastic case can be rewritten as
\begin{eqnarray*}
  \vec{r}_k^{v_i}
    &=& \frac{1}{\rho_k} \sum_{n=1}^{3} \vec{D}_n \left[\sum_{m=1}^{3} g_{k,m,n}^\text{vol} \vec{q}_k^{\sigma_{im}}\right]
    + \mat{L} \vec{p}_{k}^{v_i}, \quad \text{for } i=1,2,3, \\
  \vec{r}_k^{\sigma_{ij}}
    &=& \delta_{ij} \lambda_k \sum_{m=1}^{3} \sum_{n=1}^{3} g_{k,m,n}^\text{vol} \Big[\vec{D}_n \vec{q}_{k}^{v_m}\Big]
    + \mu_k \sum_{m=1}^3 \Big(g_{k,i,m}^\text{vol} \Big[\vec{D}_m \vec{q}_k^{v_i}\Big]
                            + g_{k,m,i}^\text{vol} \Big[\vec{D}_i \vec{q}_k^{v_m}\Big]\Big)
    + \mat{L} \vec{p}_{k}^{\sigma_{ij}}, \quad \text{for } i,j=1,2,3.
\end{eqnarray*}

The volume kernel performs element-wise tasks corresponding to the volume terms, which are the first terms of equations \eqref{eqn:dgRhsAcou1}-\eqref{eqn:dgRhsAcou2} in the acoustic case.
For each element, the values of fields are fetched from the array $\mathtt{Q}$, and the main algebraic operations are linear combinations and six matrix-vector products with squared matrices of size $N_p \times N_p$.
The surface kernel computes the penalty terms and performs the matrix-vector products for the second terms of equations \eqref{eqn:dgRhsAcou1}-\eqref{eqn:dgRhsAcou2}.
\modif{In both volume and surface kernels, direct matrix-vector products are implemented.
Let us mention that alternative DG methods with sparse elemental matrices can leverage sparse matrix-vector products (see \eg \cite{breuer2014sustained,chan2015bb}).}
The implementation of the surface kernel is trickier because computing the penalty terms requires the values of traces at face nodes corresponding to both sides of the interface (\textit{i.e.} values for the current element and its neighbors), which leads to erratic memory accesses when fetching neighbor values.
We use a mapping array that, for each face node, gives the location of the first trace corresponding to the neighbor element in the array $\mathtt{Qf}$.
For each element, the surface kernel then loads both local and neighbor values of traces from the array $\mathtt{Qf}$, computes the penalty terms and multiplies them with geometric factors.
The obtained values are then used in the merged matrix-vector product with a rectangular matrix of size $N_p \times N_\text{faces}N_{fp}$ and a vector of size $N_\text{faces}N_{fp}$.
Both volume and surface kernels accumulate the final right-hand side terms in the current right-hand side array $\mathtt{R}_0$.
The update kernel loads all the right-hand side arrays, updates the fields values in the array $\mathtt{Q}$ by performing the time-stepping, and updates the trace values in the array $\mathtt{fQ}$.

For the {one-element-per-thread} strategy, all kernels run with $256$ threads per thread block.
That number of threads has been suggested by Fuhry \etal \cite{Fuhry2014}, and is also optimal for our implementation.
For both volume and surface kernels, the elemental matrices (differentiation and lift) are stored in global memory and prefetched into shared memory by all the threads in collaborative way (successive threads fetch successive values).
Let us mention that, in the implementation of Fuhry \etal \cite{Fuhry2014}, the elemental matrices are stored in constant memory, and each value is fetched during the matrix-vector products.
This strategy was however less efficient in our case.
The geometric factors, physical parameters and fields/traces, which are local to each element, are stored in registers.
The final values of the right-hand side terms are stored in temporary arrays before being transferred to local memory using coalesced data transfers.
While this implementation uses a lot of registers, leading to spilling registers for high polynomial degree, it provides the best runtime with this strategy.
The kinds of memory transfer are summarized in Table \ref{tab:arraysTransfer}.

With the {one-node-per-thread} strategy, threads collaborate to perform the computations required for each element.
For all the kernels, each thread block deals with several elements, which allow to have a reasonable number of threads per thread block for small polynomial degree.
Therefore, we have $K_\text{blkV} N_p$ threads per thread block for the volume kernel, $K_\text{blkS}\:\text{max}(N_p,N_\text{faces}N_{fp})$ for the surface kernel, and $K_\text{blkU} N_p$ for the update kernel.
The numbers of elements per thread block, $K_\text{blkV}$, $K_\text{blkS}$ and $K_\text{blkU}$, must be tuned.
This implementation make use of shared memory to store temporary arrays with values that are useful for different threads.
For more details about this implementation, we refer to our previous paper \cite{Modave2015}.

\begin{table}
\centering
\begin{tabular}{l|l|l} \hline
\multirow{2}{*}{Name} & Memory transfer & Memory transfer \\
     & \textit{one-element-per-thread} & \textit{one-node-per-thread} \\ \hline
Fields and traces &
    Fetch into registers &
    Fetch into registers \\
Temporary array &
    Store in registers &
    Store in shared memory \\
Geometric and physical factors & 
    Prefetch into registers & 
    Prefetch into shared memory \\
Elemental matrices &
    Prefetch into shared memory &
    Fetch into registers \\ \hline
\end{tabular}
\caption{Memory transfers performed by the volume and surface kernels.
Fetched data are transferred when required, while prefetched data are loaded at the beginning of each kernel execution.
The temporary array stores the vectors to be used in the matrix-vector products, which are the penalty vectors for the surface kernel.}
\label{tab:arraysTransfer}
\end{table}

%% file: implementCUBLAS.tex
\subsection{Implementations with SGEMM routine}
\label{sec:imp:CUBLAS}

We investigate an alternative implementation which makes use of an extern BLAS library.
Such library provides linear algebra routines that are tuned for specific hardware devices.
In our computational procedure, the matrix-vector products of equations \eqref{eqn:dgRhsAcou1}-\eqref{eqn:dgRhsAcou2} can be merged into global matrix-matrix products, which are BLAS-3 operations.
They will be performed here with the SGEMM routine of the cuBLAS library developed by Nvidia \cite{cublas}.
We describe hereafter the procedure only for the acoustic case, which can be straightforwardly adapted to the elastic case.

In order to integrate global matrix-matrix products in the computational procedure, we derive an alternative partition of tasks.
The computation of the right-hand side terms can be rewritten with global operations.
By merging equations \eqref{eqn:dgRhsAcou1}-\eqref{eqn:dgRhsAcou2} for all the elements, we have
\begin{eqnarray}
  \mat{R}^p
    &=& \sum_i \vec{D}_i\mathcal{L}^{p}_i\!\left(\mat{Q}^{\vec{v}}\right)
    + \mat{L} \mat{P}^p, \label{eqn:dgRhsGlobAcou1} \\
  \mat{R}^{\vec{v}}
    &=& \mathcal{L}^{\vec{v}}\!\left(\vec{D}_1 \mat{Q}^p, \mat{D}_2 \mat{Q}^p, \mat{D}_3 \mat{Q}^p\right)
    + \mat{L} \mat{P}^{\vec{v}}, \label{eqn:dgRhsGlobAcou2}
\end{eqnarray}
where $\mat{Q}^{\{p,\vec{v}\}}$, $\mat{P}^{\{p,\vec{v}\}}$ and $\mat{R}^{\{p,\vec{v}\}}$ are matrices for which the $k^\text{th}$ column corresponds respectively to the values of fields, the penalty terms and the right-hand side vectors of element $k$.
The operators $\mathcal{L}^{p}_i$ and $\mathcal{L}^{\vec{v}}$ abstract the linear combinations to perform for the volume terms.
A rapid study of equations \eqref{eqn:dgRhsGlobAcou1}-\eqref{eqn:dgRhsGlobAcou2} leads to the new partition of tasks:
\begin{enumerate}
  \item computation of the \textit{gradient} matrix-matrix product for equation \eqref{eqn:dgRhsGlobAcou2} \textit{[SGEMM routine]}:
    \begin{equation}
      \begin{bmatrix} \mathtt{D}_1 \\ \mathtt{D}_2 \\ \mathtt{D}_3 \end{bmatrix}
      \mathtt{Q}^p
      \quad\rightarrow\quad
      \begin{bmatrix}  \mathtt{Q}^\text{tmp}_1 \\ \mathtt{Q}^\text{tmp}_2 \\ \mathtt{Q}^\text{tmp}_3 \end{bmatrix},
      \quad\quad\quad
      \text{\color{gray}(or $\mathtt{D}_\mathrm{grad} \mathtt{Q}^p \ \rightarrow \ \mathtt{Q}^\mathrm{tmp}$)}
      \label{eqn:cublasGemm1}
    \end{equation}
  \item computation of the linear combinations for the volume terms \textit{[volume kernel]}:
    \begin{eqnarray*}
      \mathcal{L}^{\vec{v}}\!\left(\mathtt{Q}^\text{tmp}_1, \mathtt{Q}^\text{tmp}_2, \mathtt{Q}^\text{tmp}_3\right)
        &\ \rightarrow\ & \mathtt{R}^{\vec{v}}, \\
      \mathcal{L}^{p}_i\!\left(\mathtt{Q}^{\vec{v}}\right)
        &\ \rightarrow\ & \mathtt{Q}^\text{tmp}_i,
        \quad\quad \text{for } i=1,2,3,
    \end{eqnarray*}
  \item computation of the \textit{divergence} matrix-matrix product for equation \eqref{eqn:dgRhsGlobAcou1} \textit{[SGEMM routine]}:
    \begin{equation}
      \begin{bmatrix} \mathtt{D}_1 & \mathtt{D}_2 & \mathtt{D}_3 \end{bmatrix}
      \begin{bmatrix} \mathtt{Q}^\text{tmp}_1 \\ \mathtt{Q}^\text{tmp}_2 \\ \mathtt{Q}^\text{tmp}_3 \end{bmatrix}
      \quad\rightarrow\quad
      \mathtt{R}^p,
      \quad\quad
      \text{\color{gray}(or $\mathtt{D}_\text{div} \mathtt{Q}^\text{tmp} \ \rightarrow \ \mathtt{R}^p$)}
      \label{eqn:cublasGemm2}
    \end{equation}
  \item computation of penalty vectors and storage in arrays $\mathtt{P}^p$ and $\mathtt{P}^{v_i}$ \textit{[surface kernel]},
  \item computation of the \textit{lift} matrix-matrix products and addition to the right-hand side arrays \textit{[2 $\times$ SGEMM routine]}:
    \begin{eqnarray*}
      \mathtt{L} \mathtt{P}^p + \mathtt{R}^p
        &\ \rightarrow\ & \mathtt{R}^p, \\
      \mathtt{L} \mathtt{P}^{\vec{v}} + \mathtt{R}^{\vec{v}}
        &\ \rightarrow\ & \mathtt{R}^{\vec{v}},
    \end{eqnarray*}
  \item update of fields and traces with the time stepping scheme \textit{[update kernel]}.
\end{enumerate}
In this new procedure, the first three operations compute the volume terms, the next two compute the surface terms, and the latter performs the update.
The procedure requires three kernels and four calls of the SGEMM routine.
We describe hereafter the arrays that must stored in global memory, the different settings for the SGEMM routine, and the kernels.

In comparison with the previous implementations, arrays $\mathtt{Q}^\text{tmp}_i$ and $\mathtt{P}^{\{p,\vec{v}\}}$ have been introduced to store intermediate results, which was not necessary before since these intermediate results were immediately used to compute matrix-vector products.
For the sake of clarity, the fields, traces and right-hand side terms are stored in two set of arrays, one for the pressure ($\mathtt{Q}^p$, $\mathtt{Qf}^p$ and $\mathtt{R}^p$) and the other for the velocity ($\mathtt{Q}^{\vec{v}}$, $\mathtt{Qf}^{\vec{v}}$ and $\mathtt{R}^{\vec{v}}$).
To efficiently use the SGEMM routine, the arrays $\mathtt{Q}^\text{tmp}_i$ are merged into the array $\mathtt{Q}^\text{tmp}$, and the differentiations arrays $\mathtt{D}_i$ are stored into two merged arrays, the gradient array $\mathtt{D}_\mathrm{grad}$ and the divergence array $\mathtt{D}_\mathrm{div}$.
The granularity of storage follows the {one-node-per-thread} strategy of the previous section, with the node index $n$ at the finest level.
This is required to use the SGEMM routine.

Conforming to BLAS terminology, the SGEMM routine computes the general matrix multiplication $\mat{C} = \alpha \mat{A} \mat{B} + \beta \mat{C}$, where $\mat{A}$, $\mat{B}$ and $\mat{C}$ are respectively $m\times k$, $k\times n$ and $m\times n$ matrices and $\alpha$ and $\beta$ are scalars.
The memory storage of arrays must be column-major for the matrix $\mat{C}$, and either column-major or row-major for the others.
They can be padded in order to improve computational performances.
The padding for DG operations is studied in section \ref{sec:perf:gemm:padd}.

The kernels of this implementation, corresponding to steps 2, 4 and 6, perform only data transfers and algebraic operations that are local to nodes or face nodes.
The volume and update kernels are implemented following the {one-node-per-thread} strategy, with $K_\text{blkV}$ and $K_\text{blkU}$ elements per thread block.
In contrast with the previous implementations, the surface kernel iterates over all the interfaces of the mesh.
Each thread computes the numerical fluxes corresponding to one face node for both sides of the interface, and each thread block deals with $K_\text{blkS}$ interfaces.

%% file: perfGemmPadd.tex
\subsection{Storage strategy for SGEMM routine}
\label{sec:perf:gemm:padd}

The performance of the SGEMM routine strongly depends on both the dimensions of matrices and their storage in memory.
For the operations required by the DG implementation, better performances are obtained with a row-major storage for $\mat{A}$, while a column-major storage is used for $\mat{B}$ and $\mat{C}$.
We have tested several padding strategies.
Only a padding of both the leading storage dimension $ldc$ and the matrix dimension $m$, with $ldc=m$, can significantly improve performances.
Increasing $ldc$ involves a larger memory storage for matrix $\mat{C}$, while increasing $m$ also adds dummy floating-point operations.
Though these tunings artificially increase the required computational resources, they lead to better runtimes for the SGEMM routine.
The dimensions of matrices and storing arrays for the different SGEMM calls are summarized in Table \ref{tab:cublas:dim}.

The matrix $\mat{C}$ corresponds to different arrays in our DG implementation: $\mathtt{Q}^\mathrm{tmp}$ for the gradient product, $\mathtt{R}^p$ for the divergence product, and $\mathtt{R}^p$ and $\mathtt{R}^{\vec{v}}$ for the lift products.
The value to pad is $3N_p$ for the temporary array $\mathtt{Q}^\mathrm{tmp}$, and $N_p$ for the right-hand side arrays $\mathtt{R}^p$ and $\mathtt{R}^{\vec{v}}$.
For the sake of consistency, the padding used for the right-hand side arrays is also used for the arrays $\mathtt{Q}^p$ and $\mathtt{Q}^{\vec{v}}$, which have the same kind of memory storage.
Since $\mathtt{Q}^p$ and $\mathtt{Q}^\mathrm{tmp}$ correspond to the matrix $\mat{B}$ of  operations gradient and divergence, respectively, the parameter $ldb$ of these operations has to be padded (see Table \ref{tab:cublas:dim}).
This additional padding does not change the performance, and is made only for compatibility.

\begin{table}[!tb]
\begin{center}
\begin{tabular}{|l|ccc|ccc|} \hline
    & $m$ & $n$ & $k$ & $lda$ & $ldb$ & $ldc$ \\ \hline
  Gradient
    & $(3N_p)_\text{pad}$ & $K$ & $N_p$
    & $N_p$ & $(N_p)_\text{pad}$ & $(3N_p)_\text{pad}$ \\
  Divergence
    & $(N_p)_\text{pad}$  & $K$ & $3N_p$
    & $3N_p$ & $(3N_p)_\text{pad}$ & $(N_p)_\text{pad}$ \\
  Lift
    & $(N_p)_\text{pad}$ & $N_{ftl}K$ & $N_\text{faces}N_{fp}$
    & $N_\text{faces}N_{fp}$ & $N_\text{faces}N_{fp}$ & $(N_p)_\text{pad}$ \\ \hline
\end{tabular}
\end{center}
\caption{Dimensions of matrices for the four SGEMM operations $\mat{C} = \alpha\mat{A}\mat{B} + \beta\mat{C}$.
  According to BLAS terminology, the dimensions of $\mat{A}$, $\mat{B}$ and $\mat{C}$ are respectively $m\times k$, $k\times n$ and $m\times n$, and $\alpha$ and $\beta$ are scalars.
  Storage is row-major for $\mat{A}$ and column-major for both $\mat{B}$ and $\mat{C}$.
  $lda$, $ldb$ and $ldc$ are the leading dimensions of storing arrays.
  $N_p$, $N_{fp}$, $N_\text{faces}$ and $K$ are respectively the number of nodes per element, the number of nodes per face, the number of faces per element, and the number of elements in the mesh.
  For the lift operation, $N_{ftl}$ is the number of fields to lift (1 for $p$ and 3 for $\vec{v}$).}
\label{tab:cublas:dim}
\end{table}

\begin{table}[!tb]
\begin{center}
\begin{tabular}{|c|cc|cc|} \hline
  $N$ & $N_p$ & $(N_p)_\text{pad}$
      & $3N_p$ & $(3N_p)_\text{pad}$ \\ \hline
  1 &   4 &   4 &  12 &  16 \\
  2 &  10 &  10 &  30 &  32 \\
  3 &  20 &  20 &  60 &  64 \\
  4 &  35 &  35 & 105 & 128 \\
  5 &  56 &  56 & \cellcolor{gray!25}\!\! 168 & \cellcolor{gray!25}\!\! 256 \\
  6 &\cellcolor{gray!25}\!\! 84 &\cellcolor{gray!25}\!\! 128 & 252 & 256 \\
  7 & 120 & 128 & 360 & 384 \\
  8 &\cellcolor{gray!25}\!\! 165 &\cellcolor{gray!25}\!\! 256 & 495 & 512 \\ \hline
\end{tabular}
\end{center}
\caption{Unpadded and optimum padded values for $N_p$ and $3N_p$ when using the SGEMM routine for the DG operations.
The padded values have been obtained by minimizing the runtimes for DG settings corresponding to approx. 8 millions discrete unknowns.
Gray cells correspond to pads leading to increases of dimensions larger than $50\%$.
No padding is used for $N_p$ from the first to the fifth polynomial degree.}
\label{tab:cublas:pad}
\end{table}

The unpadded and optimized padded values for $N_p$ and $3N_p$ are listed in Table \ref{tab:cublas:pad}.
Padding $N_p$ improves the runtime only over polynomial degree $N>5$, while padding $3N_p$ is improves runtime for all degrees.
For nearly all the cases, the best padding corresponds to increasing $N_p$ and $3N_p$ until the next power of $2$.
The single exception is for $3N_p$ at the seventh degree, where the next multiple of $32$ is better.
{\color{black}For small polynomial degrees $(N\leq5)$, we have observed speedups between $1.1$ and $1.5$ for the gradient operation.
The paddings do not affect the other operations.
For higher degrees, speedups are between $2.1$ and $3.2$ for the gradient operation and between $1.4$ and $1.8$ for the others.}
In some cases (gray cells in Table \ref{tab:cublas:pad}), these paddings lead to an increase of the size of the system larger than $50\%$.


%% file: perfGemmProfile.tex
\subsection{Profiling of SGEMM routine for DG operations}
\label{sec:perf:gemm:prof}

To evaluate hardware utilization, we compare both arithmetic throughput and memory bandwidth of kernels to the theoretical peak values provided by the constructor, which are respectively 4,616\:GFLOP/s (single precision) and 224\:GB/s for Nvidia's GTX980.
The number of floating-point operations required for the general matrix multiplication is
\begin{eqnarray*}
  \#(\text{FLOP}) = 2 m n (k + 1)
\end{eqnarray*}
and the number of bytes transfered from and to the global memory is
\begin{eqnarray*}
  \#(\text{Byte}) = 4 (mk + nk + 2mn),
\end{eqnarray*}
where the dimensions $m$, $n$ and $k$ are not padded.
Dividing these values by the runtime of the routine provides  \textit{estimated arithmetic throughput} and  \textit{estimated memory bandwidth}, respectively.
These metrics give information about the performances for the requested operations, but they assume that there are no dummy floating-point operations and that data is always useful and transferred only once.
However, optimization strategies, such as padding, can artificially increase the work really performed by the device.
To evaluate that work, we also consider the \textit{effective arithmetic throughput} and the \textit{effective memory bandwidth}, which are obtained using metrics provided by Nvidia's profiler \texttt{nvprof}.
The effective arithmetic throughput is computed using the number of floating-point operations given by \verb!flop_count_sp!, and the effective memory bandwidth is obtained by summing \verb!dram_read_throughput! and \verb!dram_write_throughput!.

\begin{figure}[!t]
\centering
\begin{tabular}{cc}
  \begin{subfigure}[b]{8cm}
    \centering
    \caption{Estimated and effective throughputs} \vspace{-0.6cm}
    \includegraphics[width=7.5cm]{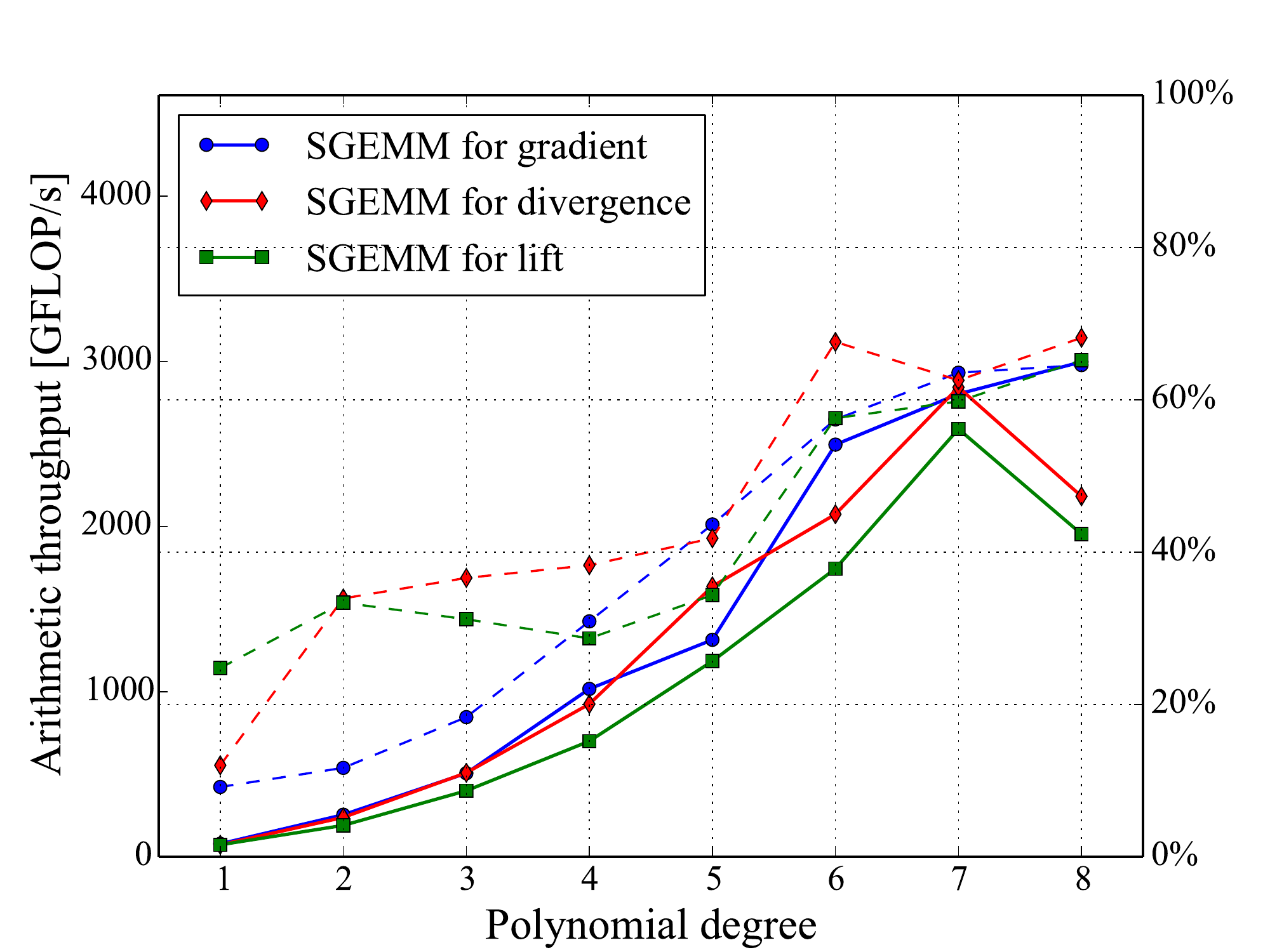}
    \label{fig:perfGEMM:GFLOPs}
  \end{subfigure} &
  \begin{subfigure}[b]{8cm}
    \centering
    \caption{Estimated and effective bandwidths} \vspace{-0.6cm}
    \includegraphics[width=7.5cm]{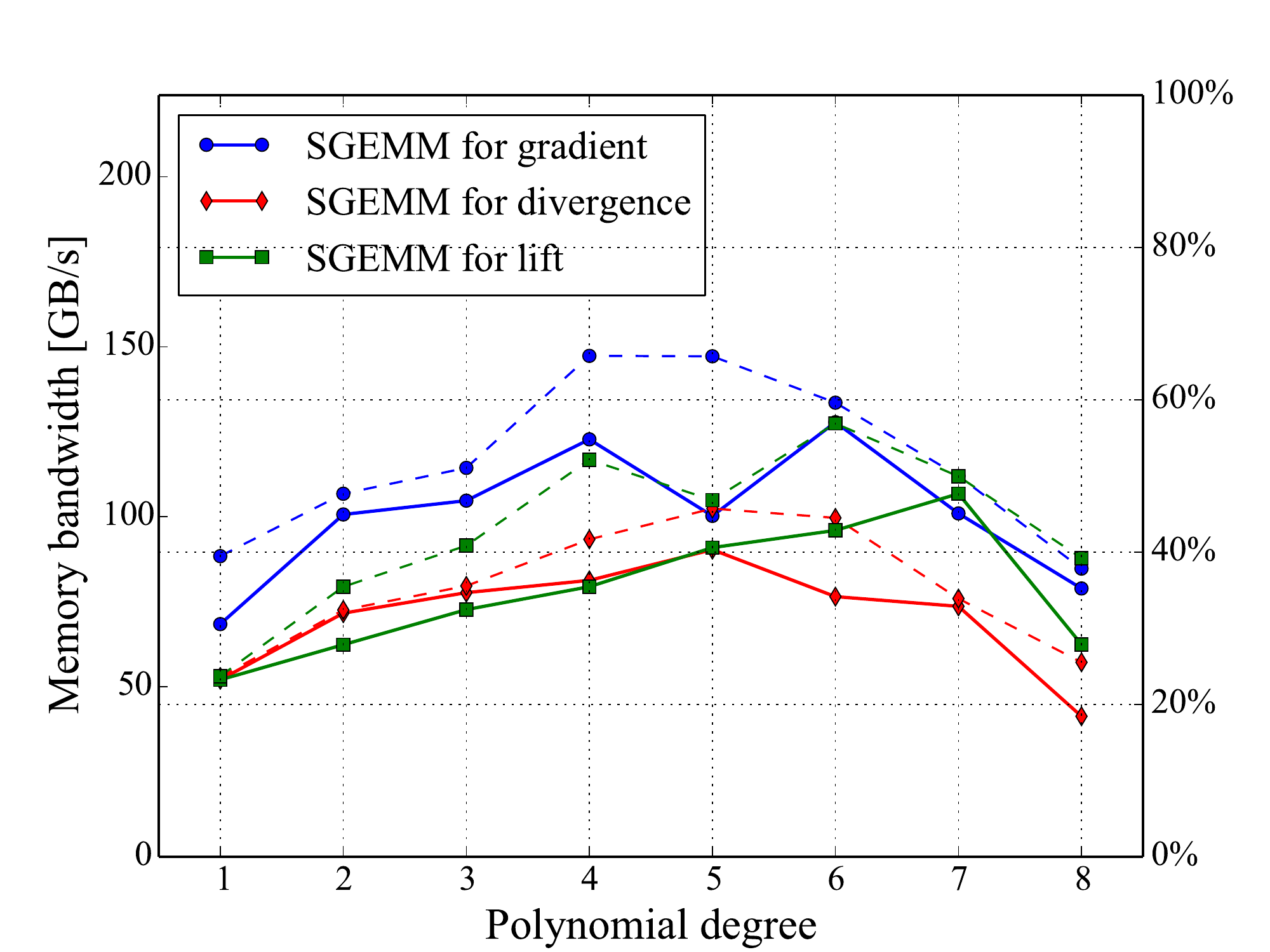}
    \label{fig:perfGEMM:BW}
  \end{subfigure}
\end{tabular}
\caption{Arithmetic throughput $(a)$ and memory bandwidth $(b)$ for the SGEMM routine for DG operations with several polynomial degrees \modif{and, for each degree, $N_p K \approx $ 2 millions}.
Continuous lines correspond to estimated performances (\textit{i.e.} considering only requested operations), while discontinuous lines are for effective performances (\textit{i.e.} considering all the operations achieved by the device).}
\label{fig:perfGEMM}
\end{figure}

Figure \ref{fig:perfGEMM} shows the estimated and effective performances of cuBLAS SGEMM routine for the DG operations presented in section \ref{sec:imp:CUBLAS}.
Since the performance of the lift operation is similar for pressure and velocity fields, we show the results only for the pressure field.
For all the operations, the estimated arithmetic throughput is far smaller than the effective value for low degrees, especially for divergence and lift operations, while $m$ is not padded.
Large differences are also observed at high degrees when the padding significantly increases $m$: for the gradient operation with $N=5$ and for both divergence and lift operations with $N=6$ and $8$.
In particular, the estimated throughput significantly decreases from $N=7$ to $N=8$ for both divergence and lift operations, while the effective throughput slightly increases.
This difference of behavior is clearly related to the large number of dummy operations introduced by the large padding required for $N=8$, while the padded and unpadded dimensions are very close for $N=7$.
Similar differences between estimated and effective memory bandwidths are also observed for cases with large padding.

In the remainder, we consider only estimated arithmetic throughput and memory bandwidth.
Though these values underestimate the real work of the device, they allow us to compare different implementations without counting potential dummy operations and useless data transfers.

For all operations, the estimated arithmetic throughput is very small for low degrees (from few percents to $20\%$ of the peak performance between $N=1$ and $4$) and grows until approximately $60\%$ for $N=7$.
The largest throughput, $65\%$, is obtained with the gradient operation for $N=8$.
The estimated memory bandwidth increases for lower degrees and decreasing for higher degrees.
The bandwidth reaches $54-57\%$ of the peak performance (for $N=4$ and $N=6$) for the gradient operation, while the other operations exhibit slightly worse performances.

\begin{figure}[!t]
\centering
\begin{tabular}{cc}
  \begin{subfigure}[b]{8cm}
    \centering
    \caption{Roofline analysis} \vspace{-0.6cm}
    \includegraphics[width=7.5cm]{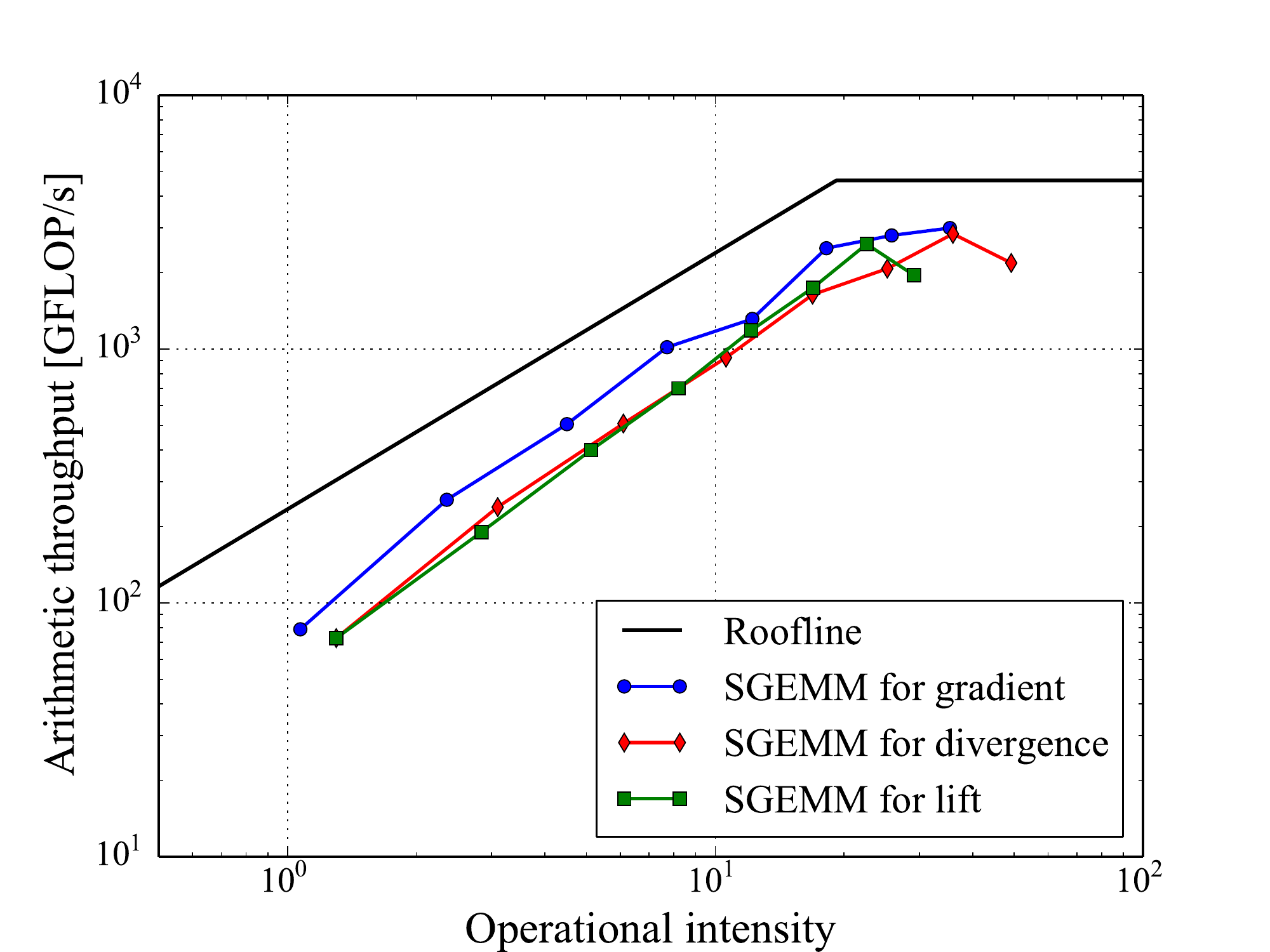}
    \label{fig:GEMM:roofline}
  \end{subfigure} &
  \begin{subfigure}[b]{8cm}
    \centering
    \caption{Performance analysis} \vspace{-0.6cm}
    \includegraphics[width=7.5cm]{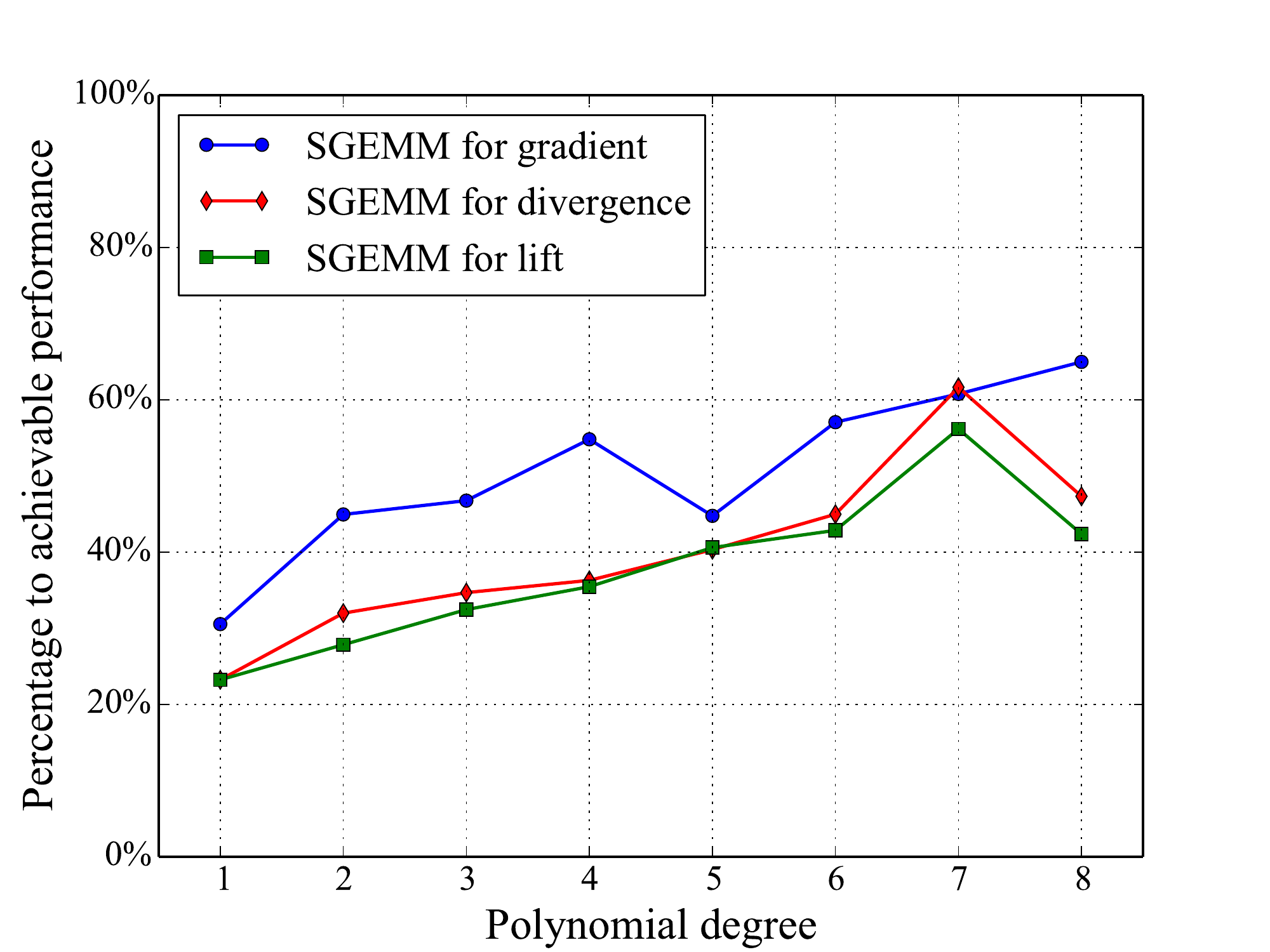}
    \label{fig:GEMM:perf}
  \end{subfigure}
\end{tabular}
\caption{Roofline analysis $(a)$ of the SGEMM routine for DG operations, and percentage to achievable performance $(b)$ considering the specific device.
For every curves, each bullet corresponds to one polynomial degree.
The leftmost bullet always corresponds to the first degree.}
\label{fig:perfGEMM:roofline}
\end{figure}

The general behavior of both throughput and bandwidth can be explained using to the roofline model.
Because of the physical limitations of the device, the performance of any operation is theoretically bounded either by the throughput or by the bandwidth, but rarely by both together.
The roofline model provides a tool to evaluate the dominant bound, and how far we are from that bound \cite{williams2009roofline}.
The arithmetic throughput is plotted as a function of the operational intensity, which is defined as
\begin{eqnarray*}
  \text{operational intensity}
    = \frac{\#(\text{FLOP})}{\#(\text{Byte})}
    = \frac{2 m n (k + 1)}{4 (mk + nk + 2mn)}.
\end{eqnarray*}
This metric determines the theoretical achievable throughput, which is
\begin{eqnarray*}
  \text{achievable throughput}
    = \min\big\{\text{peak bandwidth}\times\text{operational intensity},\ \text{peak throughput}\big\}
\end{eqnarray*}
The operation is bandwidth or throughput bounded if the operational intensity is small or large, respectively.
Figure \ref{fig:GEMM:roofline} shows the curve of the achievable throughput (called the \textit{roofline}) together with the obtained throughput for the different DG operations.
The performance of the SGEMM routine for the DG operations are theoretically bounded by the bandwidth until the sixth or seventh degree, and by the arithmetic throughput beyond these polynomial degrees.
This explains the low throughput observed for small degrees, as well as the stagnation of the throughput and the decaying memory bandwidth observed for the highest degrees.

Arithmetic throughput and memory bandwidth are representative of the implementation efficiency only when the kernel is throughput or bandwidth bounded, respectively.
To quantify the effectiveness of the SGEMM implementation in a unique way, we consider the percentage of the obtained throughput to its achievable value.
When the kernel is bandwidth bound, this percentage actually also corresponds to the percentage of the obtained bandwidth to the peak bandwidth.
This metric, that we call \textit{percentage to achievable performance}, is shown on figure \ref{fig:GEMM:perf}.
We observe that the performance of the implementation increases with the polynomial degree, but only for the cases \modif{without} large padding already discussed above.

%% file: perfDgProfile.tex
\subsection{Comparison and analysis of DG implementations}
\label{sec:perf:DG:prof}

\begin{figure}[!t]
\centering
\begin{tabular}{cc}
  \begin{subfigure}[b]{8cm}
    \centering
    \caption{Global runtime} \vspace{-0.5cm}
    \includegraphics[width=7.5cm]{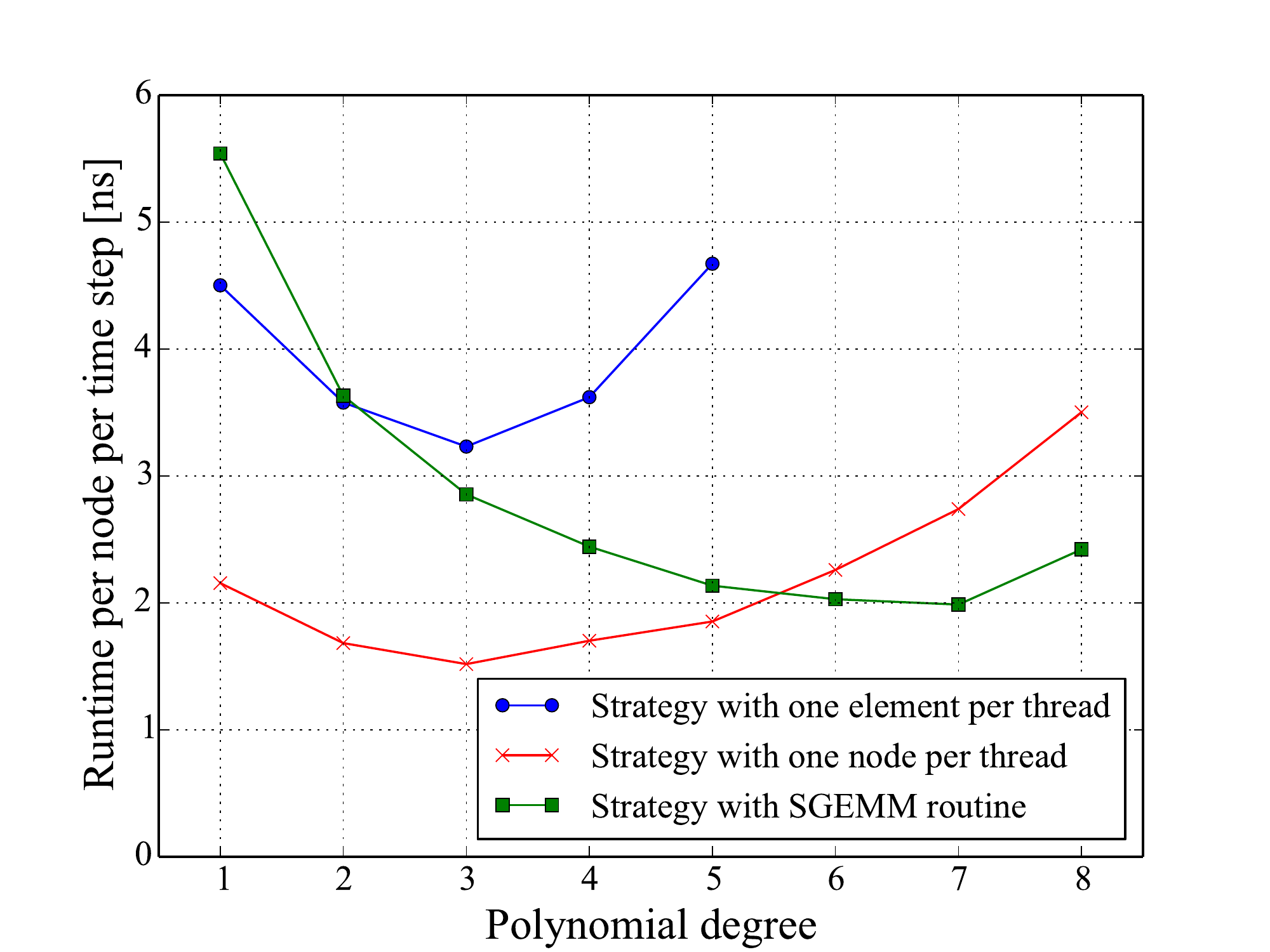}
    \label{fig:perf:runtimeAcou:global}
  \end{subfigure} &
  \begin{subfigure}[b]{8cm}
    \centering
    \caption{Runtime \it (one-element-per-thread)} \vspace{-0.5cm}
    \includegraphics[width=7.5cm]{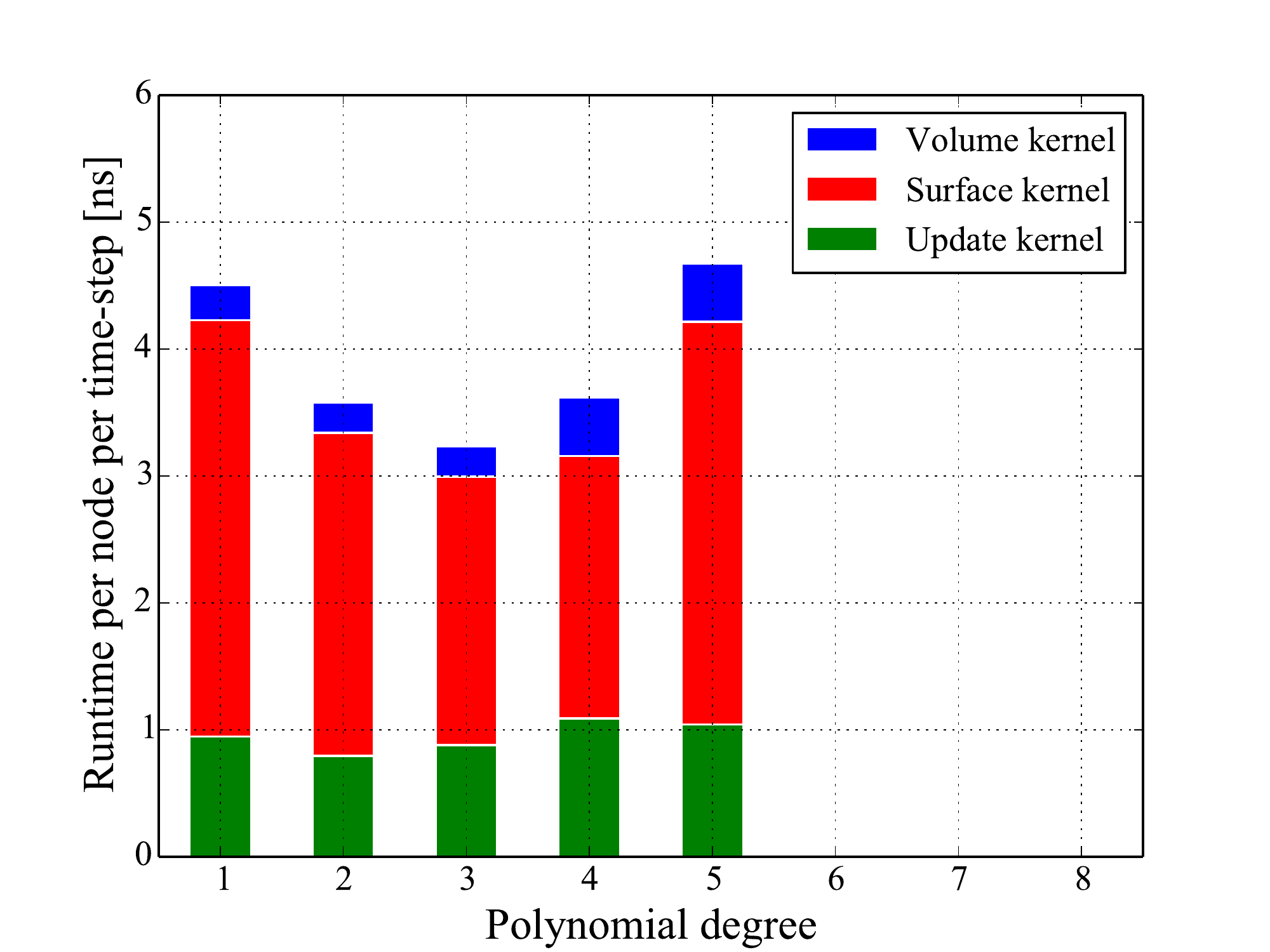}
    \label{fig:perf:runtimeAcou:EPT}
  \end{subfigure} \vspace{0.5cm} \\
  \begin{subfigure}[b]{8cm}
    \centering
    \caption{Runtime \it (one-node-per-thread)} \vspace{-0.5cm}
    \includegraphics[width=7.5cm]{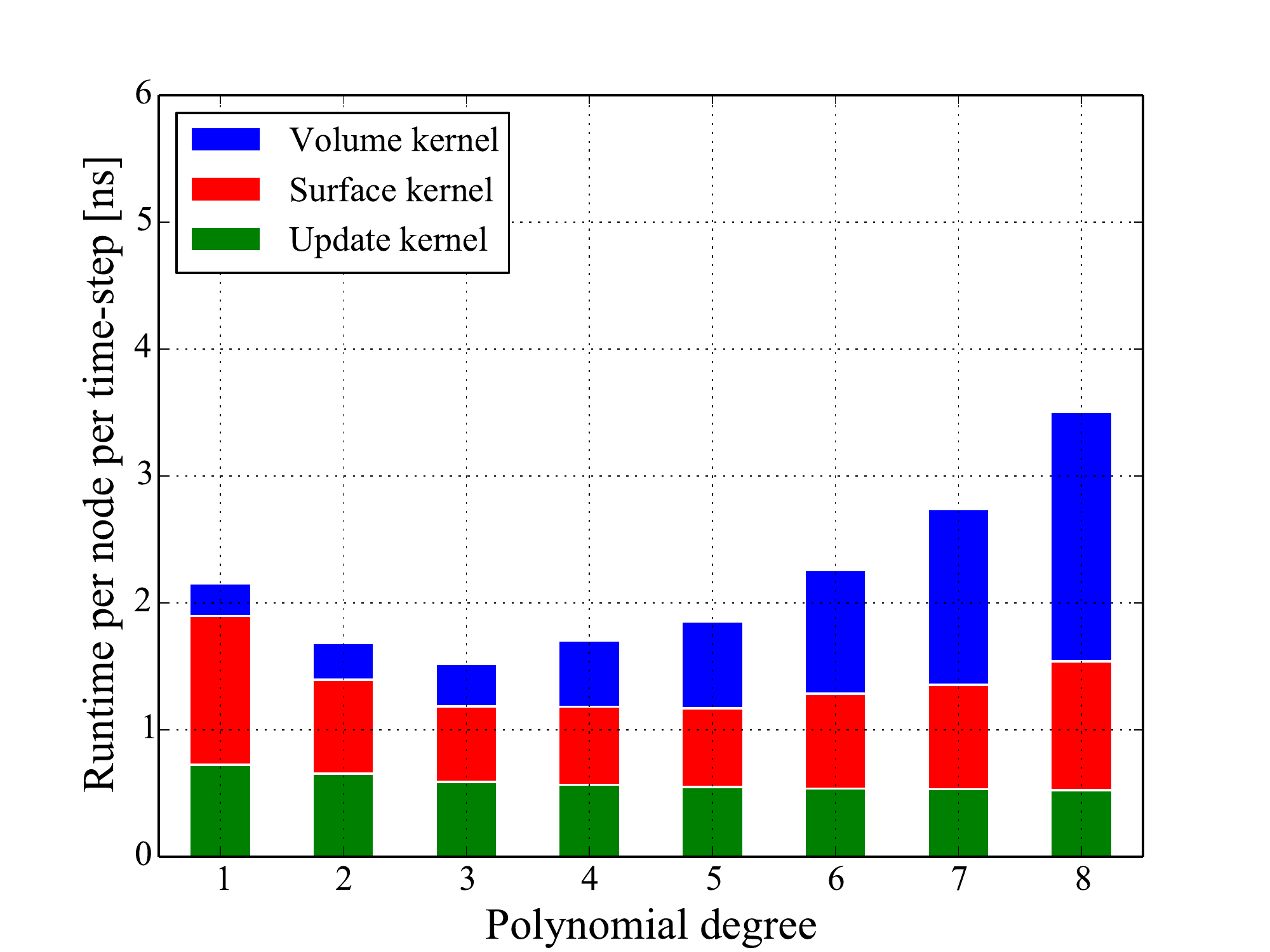}
    \label{fig:perf:runtimeAcou:NPT}
  \end{subfigure} &
  \begin{subfigure}[b]{8cm}
    \centering
    \caption{Runtime \it (strategy with SGEMM)} \vspace{-0.5cm}
    \includegraphics[width=7.5cm]{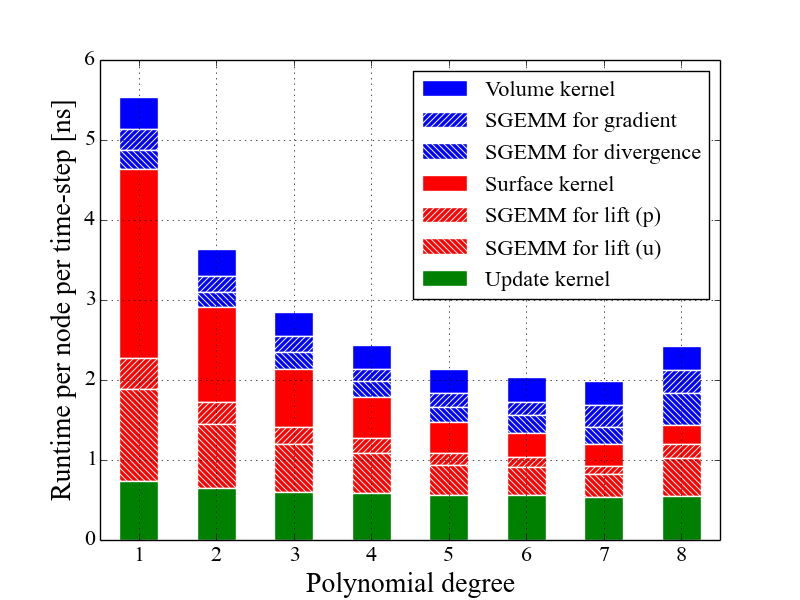}
    \label{fig:perf:runtimeAcou:Cublas}
  \end{subfigure}
\end{tabular}
\caption{Global runtimes per node per time step with the three DG implementations in the acoustic case (a), and runtimes kernel-by-kernel for each implementation (b)-(c)-(d).
\modif{For each polynomial degree, a mesh corresponding to approx. 2 millions of nodes has been used.}}
\label{fig:perf:runtimeAcou}
\end{figure}

We compare and analyse the performance of the three DG implementations.
The global runtime per node per time step is plotted on figure \ref{fig:perf:runtimeAcou} for the acoustic case.
We also show the detail kernel-by-kernel for each implementation.
For the one-element-per-thread strategy, results are shown only up to the fifth polynomial degree.
Larger degrees were not feasible due to memory limitations.
This is discussed later.

For all implementations, the global runtime per node decreases with polynomial degree until some minimum value is achieved.
After this, the runtime increases with polynomial order.
The minimum is reached at $N=3$ for both one-element-per-thread and one-node-per-thread strategies, and $N=7$ for the strategy with the SGEMM routine.
Until the fifth degree, the one-node-per-thread strategy provides the best performance.
The runtime is larger by a factor two or more with the one-element-per-thread strategy.
For higher degrees, the strategy with SGEMM is more efficient.
{\color{black}
Let us mention that these results are not necessarily correlated to the most suitable polynomial degree for applications.
A complete performance analysis would include discussions of accuracy and time stepping.
For instance, higher polynomial degrees allow for a better representation of high frequency modes, but they also reduce the allowed time step.
Such an analysis is out of the scope of this paper.}

\begin{figure}[!pth]
\centering
\begin{tabular}{cc}
  \begin{subfigure}[b]{8cm}
    \centering
    \caption{Roofline \it (one-element-per-thread)} \vspace{-0.5cm}
    \includegraphics[width=7.5cm]{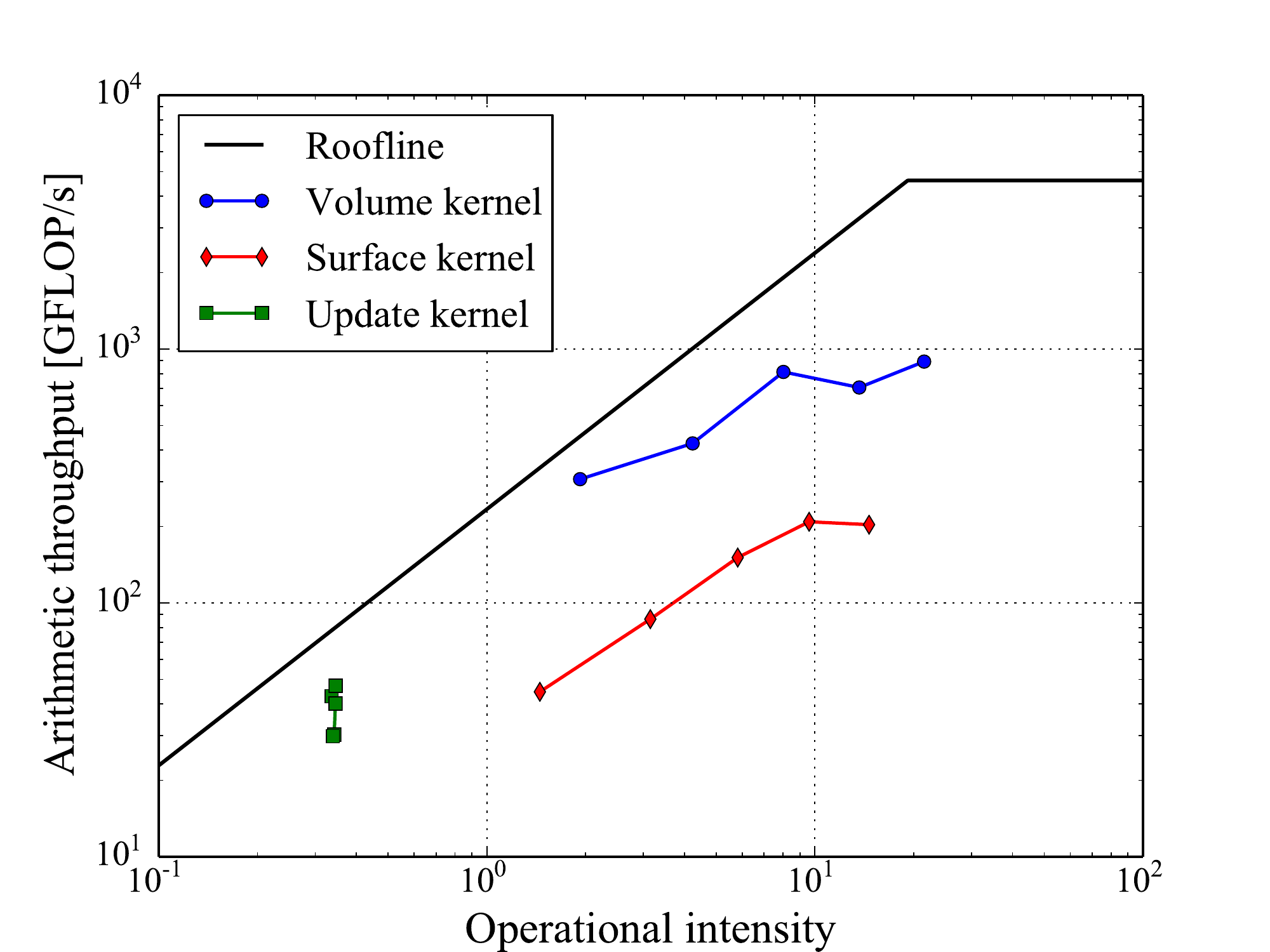}
    \label{fig:perfEPT:roofline}
  \end{subfigure} &
  \begin{subfigure}[b]{8cm}
    \centering
    \caption{Performance \it (one-element-per-thread)} \vspace{-0.5cm}
    \includegraphics[width=7.5cm]{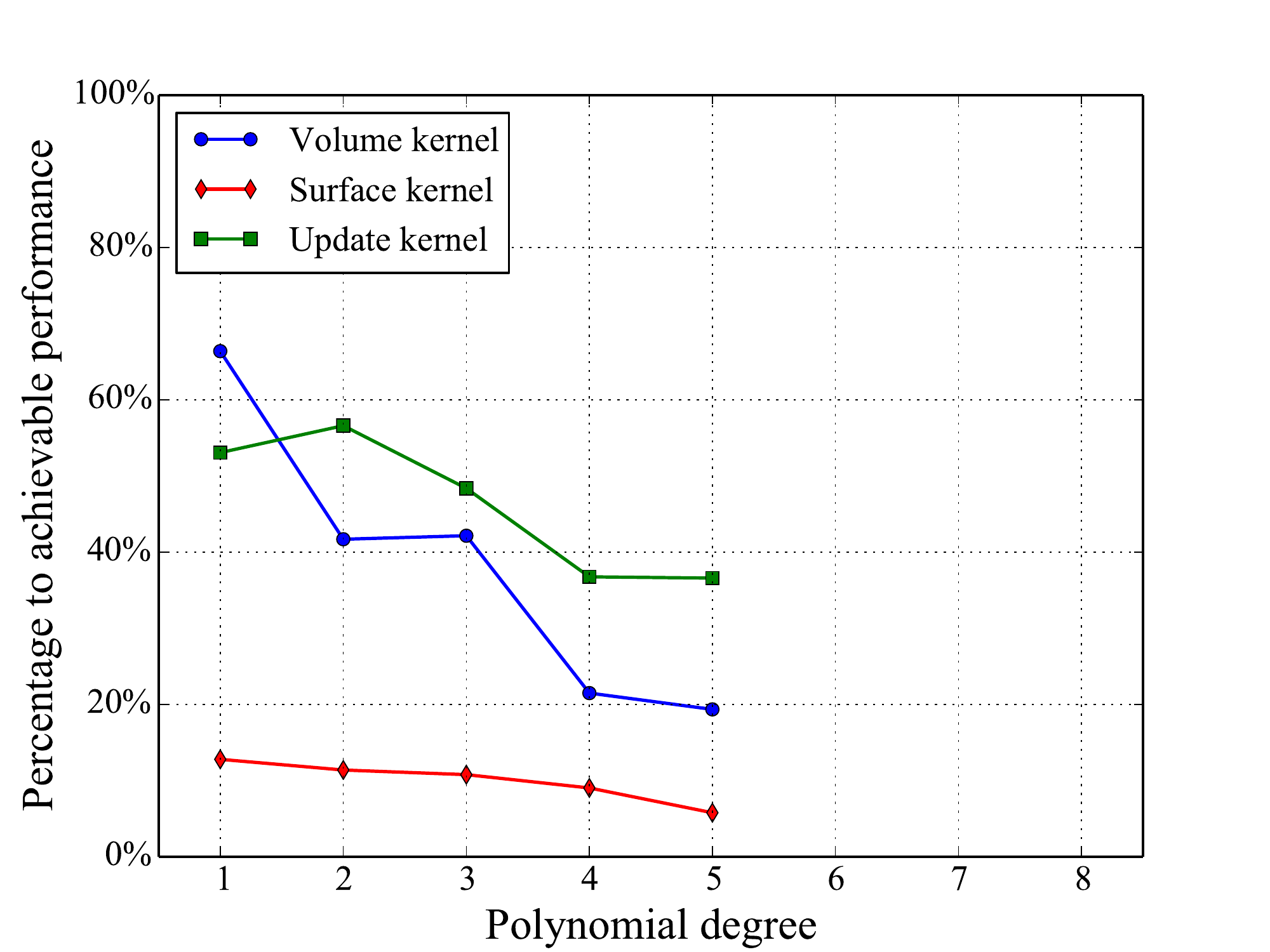}
    \label{fig:perfEPT:arch}
  \end{subfigure} \vspace{0.5cm} \\
  \begin{subfigure}[b]{8cm}
    \centering
    \caption{Roofline \it (one-node-per-thread)} \vspace{-0.5cm}
    \includegraphics[width=7.5cm]{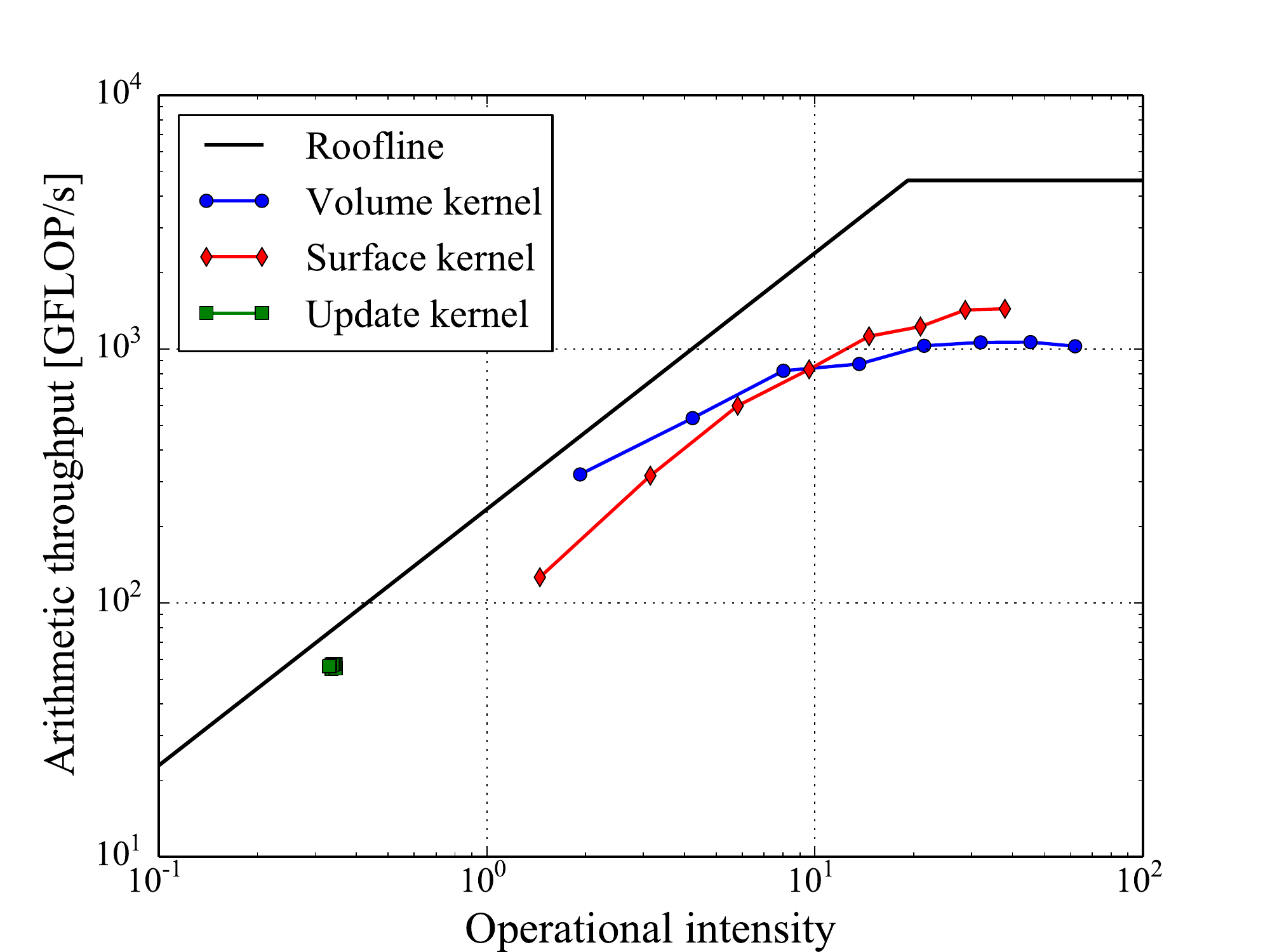}
    \label{fig:perfNPT:roofline}
  \end{subfigure} &
  \begin{subfigure}[b]{8cm}
    \centering
    \caption{Performance \it (one-node-per-thread)} \vspace{-0.5cm}
    \includegraphics[width=7.5cm]{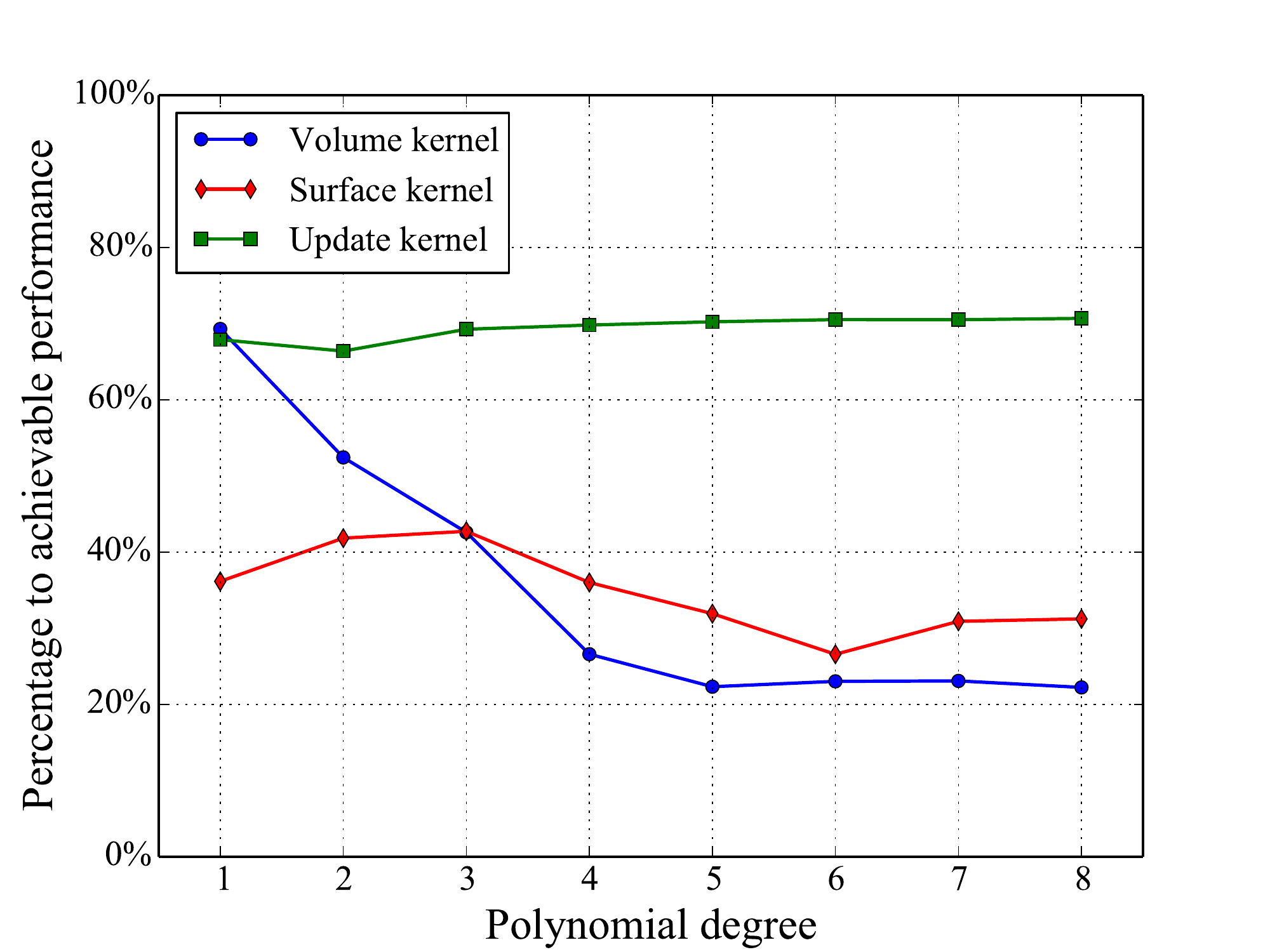}
    \label{fig:perfNPT:arch}
  \end{subfigure} \vspace{0.5cm} \\
  \begin{subfigure}[b]{8cm}
    \centering
    \caption{Roofline \it (strategy with SGEMM)} \vspace{-0.5cm}
    \includegraphics[width=7.5cm]{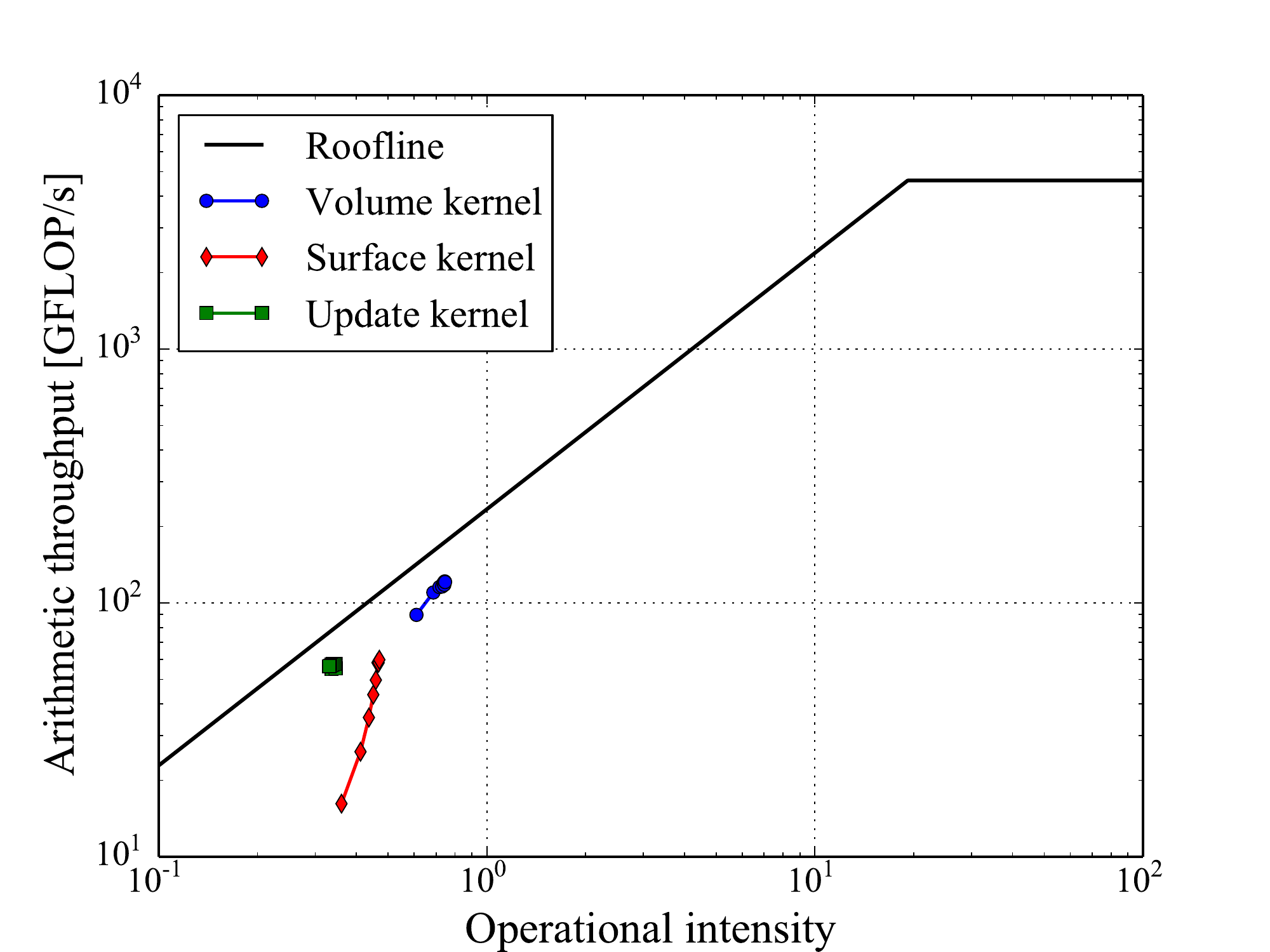}
    \label{fig:perfCublas:roofline}
  \end{subfigure} &
  \begin{subfigure}[b]{8cm}
    \centering
    \caption{Performance \it (strategy with SGEMM)} \vspace{-0.5cm}
    \includegraphics[width=7.5cm]{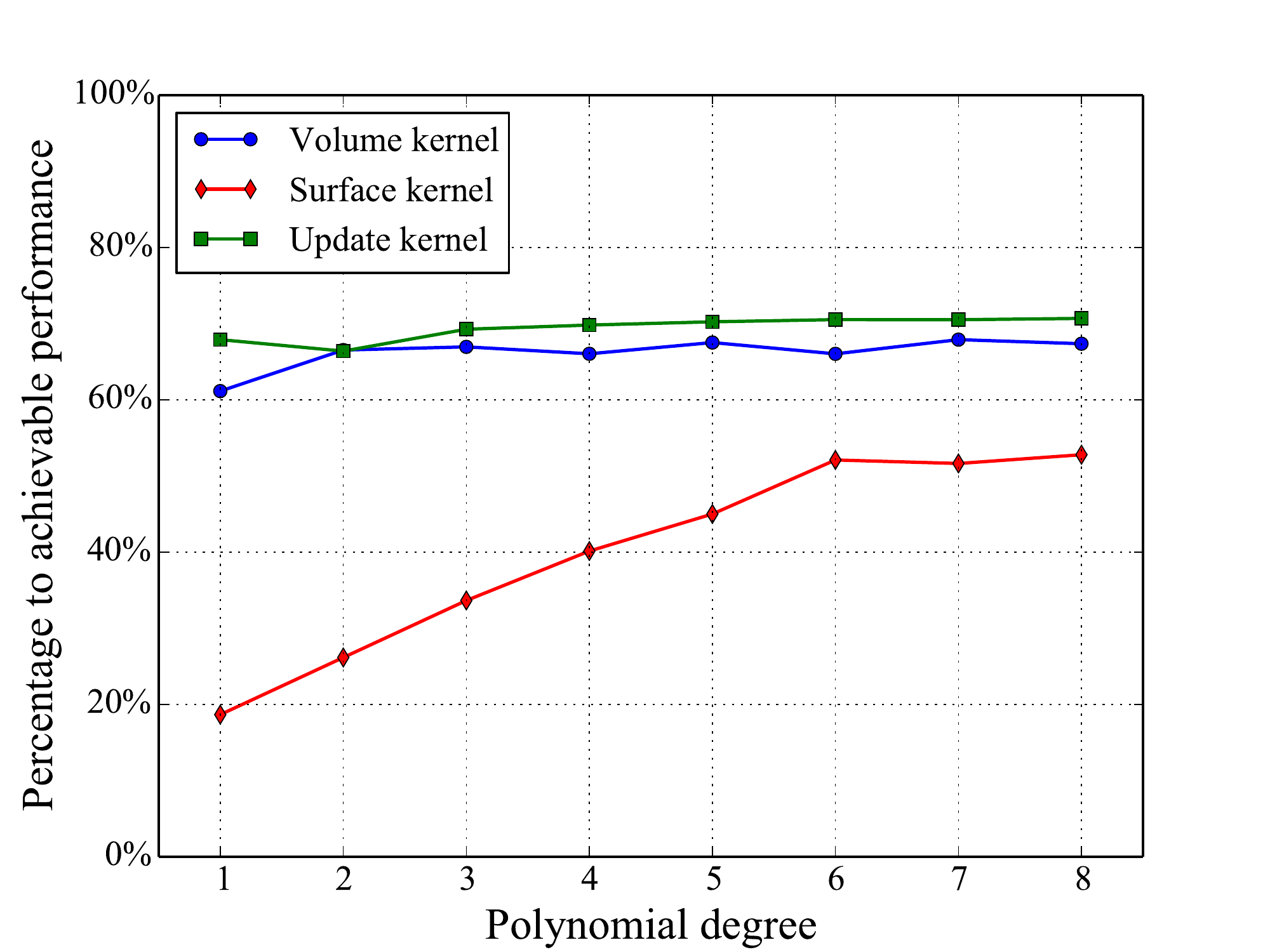}
    \label{fig:perfCublas:arch}
  \end{subfigure}
\end{tabular}
\caption{
Roofline and performance analyses of kernels for the different strategies in the acoustic case.
For every curves, each bullet corresponds to one polynomial degree.
The leftmost bullet always corresponds to the first degree.}
\label{fig:perf:roofline}
\end{figure}

In order to explain these results and to identify some bottlenecks in each implementation, we present the roofline and performance analyses of all the kernels in figure \ref{fig:perf:roofline}.
Some statistics provided by Nvidia's compiler \verb!nvcc! (allocated registers, local memory and shared memory) and Nvidia's profiler \verb!nvprof! (occupancy with \verb!achieved_occupancy! and hit rate for global loads with \verb!global_hit_rate!) are listed in table \ref{table:perf:stats}, together with the optimum numbers of elements per thread block $K_\text{blk}$.
All the kernels have been tuned and the $K_\text{blk}$'s optimized to minimize the runtime as much as possible.

\begin{table}[!pth]
\centering
\begin{tabular}{cc}
\begin{tabular}{c}
Volume kernel \textit{(one-element-per-thread)} \vspace{0.025cm} \\
\begin{tabular}{|c|c|c|c|c|c|} \hline
  $N$ & reg. & lmem &  smem & occ. & gbl.h. \\ \hline
  1 &   56 &    0 &  0.19 & 0.45 & 53\scriptsize\%\\
  2 &   64 &    0 &  1.17 & 0.35 & 51\scriptsize\%\\
  3 &  128 &    0 &  4.69 & 0.24 & 51\scriptsize\%\\
  4 &  186 &    0 & 14.35 & 0.12 & 51\scriptsize\%\\
  5 &  255 & 0.02 & 36.75 & 0.12 & 51\scriptsize\%\\ \hline
\end{tabular}
\\ \\
Surface kernel \textit{(one-element-per-thread)} \vspace{0.025cm} \\
\begin{tabular}{|c|c|c|c|c|c|} \hline
  $N$ & reg. & lmem &  smem & occ. & gbl.h. \\ \hline
  1 &  116 & 0    &  0.19 & 0.24 & 27\scriptsize\%\\
  2 &  149 & 0    &  0.94 & 0.12 & 24\scriptsize\%\\
  3 &  225 & 0    &  3.12 & 0.12 & 24\scriptsize\%\\
  4 &  255 & 0.07 &  8.20 & 0.12 & 24\scriptsize\%\\
  5 &  255 & 0.41 & 18.37 & 0.12 & 25\scriptsize\%\\ \hline
\end{tabular}
\\ \\
Update kernel \textit{(one-element-per-thread)} \vspace{0.05cm} \\
\begin{tabular}{|c|c|c|c|c|c|} \hline
  $N$ & reg. & lmem &  smem & occ. & gbl.h. \\ \hline
  1 &   72 & 0.07 & 0.05 & 0.36 & 52\scriptsize\%\\
  2 &  214 & 0.16 & 0.09 & 0.12 & 50\scriptsize\%\\
  3 &  255 & 0.38 & 0.16 & 0.12 & 50\scriptsize\%\\
  4 &   40 & 0.55 & 0.23 & 0.74 & 50\scriptsize\%\\
  5 &   40 & 0.88 & 0.33 & 0.71 & 49\scriptsize\%\\ \hline
\end{tabular}
\\ \\ \\
Volume kernel \textit{(one-node-per-thread)} \vspace{0.05cm} \\
\begin{tabular}{|c|c|c|c|c|c|} \hline
  $N$ & $K_\text{blk}$ & reg. & smem & occ. & gbl.h. \\ \hline
  1 & 16 & 32 & 1.69 & 0.93 & 76\scriptsize\% \\
  2 & 16 & 35 & 3.19 & 0.88 & 91\scriptsize\% \\
  3 & 11 & 37 & 3.91 & 0.61 & 95\scriptsize\% \\
  4 &  5 & 32 & 2.95 & 0.90 & 97\scriptsize\% \\
  5 &  9 & 37 & 8.26 & 0.73 & 95\scriptsize\% \\
  6 &  3 & 37 & 4.07 & 0.74 & 94\scriptsize\% \\
  7 &  4 & 37 & 7.67 & 0.67 & 91\scriptsize\% \\
  8 &  3 & 32 & 7.86 & 0.98 & 87\scriptsize\% \\ \hline
\end{tabular}
\\ \\
Surface kernel \textit{(one-node-per-thread)} \vspace{0.05cm} \\
\begin{tabular}{|c|c|c|c|c|c|} \hline
  $N$ & $K_\text{blk}$ & reg. & smem & occ. & gbl.h. \\ \hline
  1 & 11 & 32 & 2.15 & 0.67 & 62\scriptsize\% \\
  2 &  2 & 32 & 0.58 & 0.95 & 75\scriptsize\% \\
  3 &  2 & 32 & 0.83 & 0.92 & 80\scriptsize\% \\
  4 &  2 & 31 & 1.14 & 0.93 & 84\scriptsize\% \\
  5 &  1 & 31 & 0.76 & 0.94 & 76\scriptsize\% \\
  6 &  9 & 32 & 8.79 & 0.89 & 85\scriptsize\% \\
  7 &  7 & 32 & 8.59 & 0.88 & 83\scriptsize\% \\
  8 &  5 & 31 & 7.54 & 0.85 & 82\scriptsize\% \\ \hline
\end{tabular}
\end{tabular}
&
\begin{tabular}{c}
Volume kernel \textit{(strategy with SGEMM)} \vspace{0.05cm} \\
\begin{tabular}{|c|c|c|c|c|c|} \hline
  $N$ & $K_\text{blk}$ & reg. & smem & occ. & gbl.h. \\ \hline
  1 & 8 & 27 & 0.34 & 0.49 & 41\scriptsize\% \\
  2 & 2 & 27 & 0.09 & 0.48 & 48\scriptsize\% \\
  3 & 1 & 27 & 0.04 & 0.48 & 40\scriptsize\% \\
  4 & 1 & 27 & 0.04 & 0.91 & 44\scriptsize\% \\
  5 & 1 & 27 & 0.04 & 0.91 & 50\scriptsize\% \\
  6 & 1 & 27 & 0.04 & 0.90 & 46\scriptsize\% \\
  7 & 1 & 27 & 0.04 & 0.92 & 50\scriptsize\% \\
  8 & 1 & 27 & 0.04 & 0.85 & 50\scriptsize\% \\ \hline
\end{tabular}
\\ \\
Surface kernel \textit{(strategy with SGEMM)} \vspace{0.05cm} \\
\begin{tabular}{|c|c|c|c|c|c|} \hline
  $N$ & $K_\text{blk}$ & reg. & smem & occ. & gbl.h. \\ \hline
  1 & 8 & 26 & 0.47 & 0.50 & 59\scriptsize\% \\
  2 & 4 & 26 & 0.33 & 0.50 & 53\scriptsize\% \\
  3 & 2 & 26 & 0.23 & 0.49 & 55\scriptsize\% \\
  4 & 1 & 26 & 0.15 & 0.49 & 52\scriptsize\% \\
  5 & 1 & 26 & 0.20 & 0.49 & 63\scriptsize\% \\
  6 & 1 & 26 & 0.25 & 0.49 & 59\scriptsize\% \\
  7 & 1 & 26 & 0.32 & 0.94 & 58\scriptsize\% \\
  8 & 1 & 26 & 0.39 & 0.93 & 62\scriptsize\% \\ \hline
\end{tabular}
\\ \\
Update kernel \textit{(one-node-per-thread} \\
\textit{and strategy with SGEMM)} \vspace{0.05cm} \\
\begin{tabular}{|c|c|c|c|c|c|} \hline
  $N$ & $K_\text{blk}$ & reg. & smem & occ. & gbl.h. \\ \hline
  1 & 16 & 30 & 1.00 & 0.87 & 53\scriptsize\% \\
  2 & 16 & 32 & 2.5  & 0.88 & 48\scriptsize\% \\
  3 & 10 & 32 & 3.12 & 0.77 & 53\scriptsize\% \\
  4 &  2 & 32 & 1.09 & 0.93 & 51\scriptsize\% \\
  5 &  1 & 32 & 0.88 & 0.91 & 58\scriptsize\% \\
  6 &  1 & 32 & 1.31 & 0.92 & 55\scriptsize\% \\
  7 &  1 & 32 & 1.88 & 0.88 & 56\scriptsize\% \\
  8 &  1 & 32 & 2.58 & 0.87 & 56\scriptsize\% \\ \hline
\end{tabular}
\\ \\ \\ \\
\scriptsize
\begin{tabular}{ll}
\rowcolor{lightlightgray}
  $N$            & Polynomial degree \\
  $K_\text{blk}$ & Number of thread per thread block \\
  reg.           & Number of register per thread \\
  lmem           & Local memory allocated per thread [KB] \\
  smem           & Shared memory allocated per thread block [KB] \\
  occ.           & Occupancy \\
  gbl.h.         & Hit rate for global loads
\end{tabular}
\end{tabular}
\end{tabular}
\caption{Profiled statistics of each kernel for the different strategies in the acoustic case.
  The number of threads per thread block is always $256$ for the one-element-per-thread strategy.
  There is no local memory allocated (lmem) with the one-node-per-thread strategy and the strategy with SGEMM.}
\label{table:perf:stats}
\end{table}

The roofline analyses show that all the kernels are bandwidth bound, except those that perform matrix-vector products with a large polynomial degree.
All the update kernels and both volume and surface kernels of the strategy with SGEMM have small constant operational intensities.
These kernels mainly perform streaming operations, and require less than one floating-point operation per transferred byte.
The volume and surface kernels of both one-element-per-thread and one-node-per-thread strategies perform matrix-vector products, with which the number of required floating-point operations increases with the polynomial degree.
These kernels are in the bandwidth bound regime until $N=4$ (volume kernel) or $N=5$ (surface kernel), and in the throughput bound regime beyond.

We first analyse the performance of the one-element-per-thread kernels.
The runtime of the volume kernel is very low in comparison to the runtime corresponding to the same operation with the other strategies (blue areas on figure \ref{fig:perf:runtimeAcou}).
However, the update kernel does not perform as well as the other versions, and the surface kernel gives very bad results.
A major bottleneck of those kernels is the storage required in local and shared memory.
This storage increases with the polynomial degree $N$.
The maximum number of registers per thread for the device is not sufficient to store the local data from $N=5$ for the volume kernel, from $N=4$ for the surface kernel and for all polynomial degrees for the update kernel.
This causes register spilling where local data is stored global memory (eventually cached in L1 or L2), which increases latency.
For $N\geq6$, the shared memory required by the volume kernel is larger than the maximum allowed (48KB), and the strategy cannot be used.
The poor performance of the surface kernel is also partly explained by non-coalescing memory transfers.
With this kernel, each thread has to load data corresponding to one element and its neighbors.
Because the memory accesses are not aligned with the thread indexing for the neighbor elements, the memory transfers cannot coalesce.
The hit rate for global loads, 24\%--27\% (see table \ref{table:perf:stats}), is far smaller than for the other kernels.
We have observed that this rate rises until approx. 50\% if the memory transfers corresponding to the neighbor elements are removed.

\begin{figure}[!t]
\centering
\begin{tabular}{cc}
  \begin{subfigure}[b]{8cm}
    \centering
    \caption{Global runtime} \vspace{-0.5cm}
    \includegraphics[width=7.5cm]{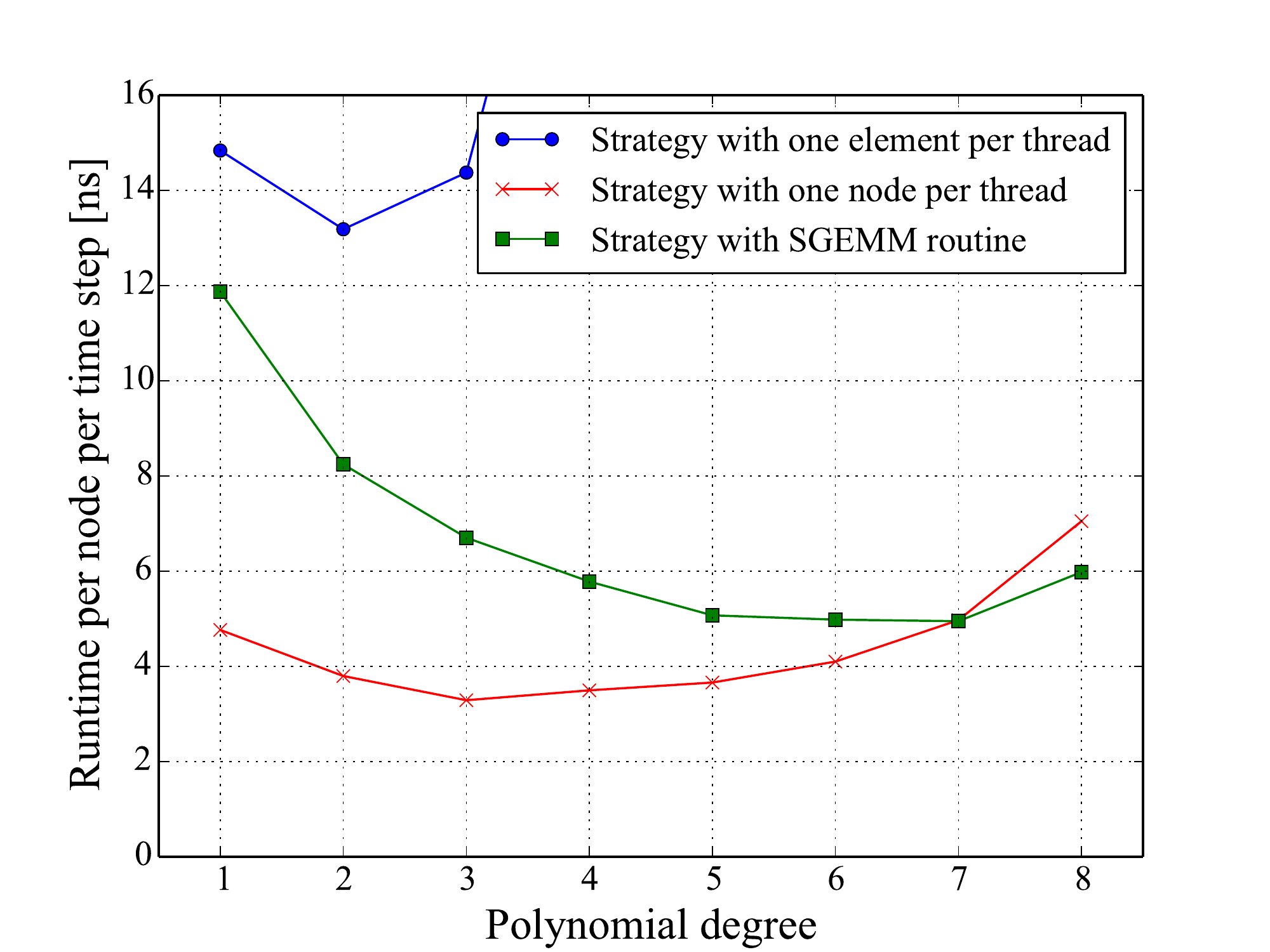}
    \label{fig:perf:runtimeElast:global}
  \end{subfigure} &
  \begin{subfigure}[b]{8cm}
    \centering
    \caption{Runtime \it (one-element-per-thread)} \vspace{-0.5cm}
    \includegraphics[width=7.5cm]{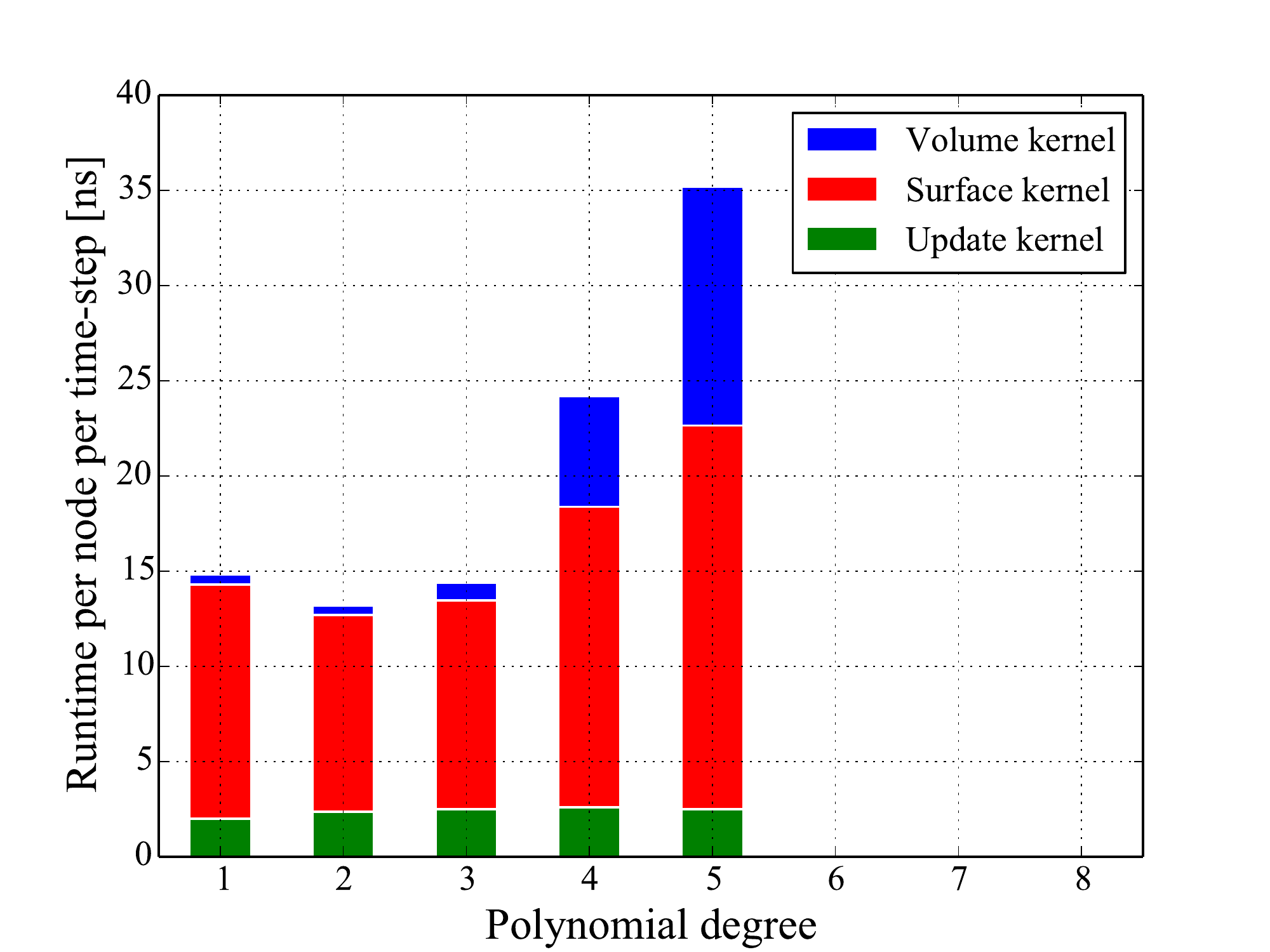}
    \label{fig:perf:runtimeElast:EPT}
  \end{subfigure} \vspace{0.5cm} \\
  \begin{subfigure}[b]{8cm}
    \centering
    \caption{Runtime \it (one-node-per-thread)} \vspace{-0.5cm}
    \includegraphics[width=7.5cm]{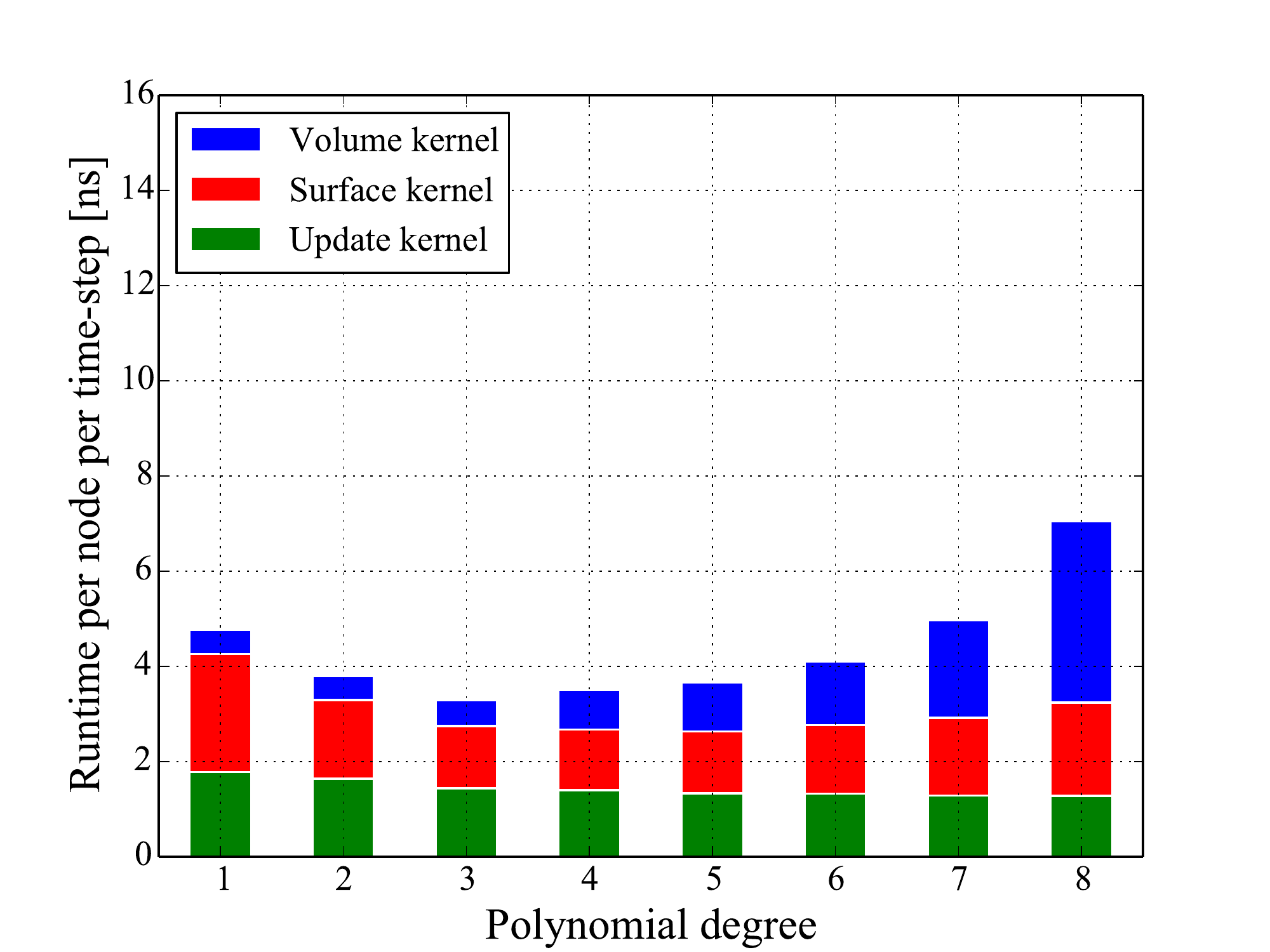}
    \label{fig:perf:runtimeElast:NPT}
  \end{subfigure} &
  \begin{subfigure}[b]{8cm}
    \centering
    \caption{Runtime \it (strategy with SGEMM)} \vspace{-0.5cm}
    \includegraphics[width=7.5cm]{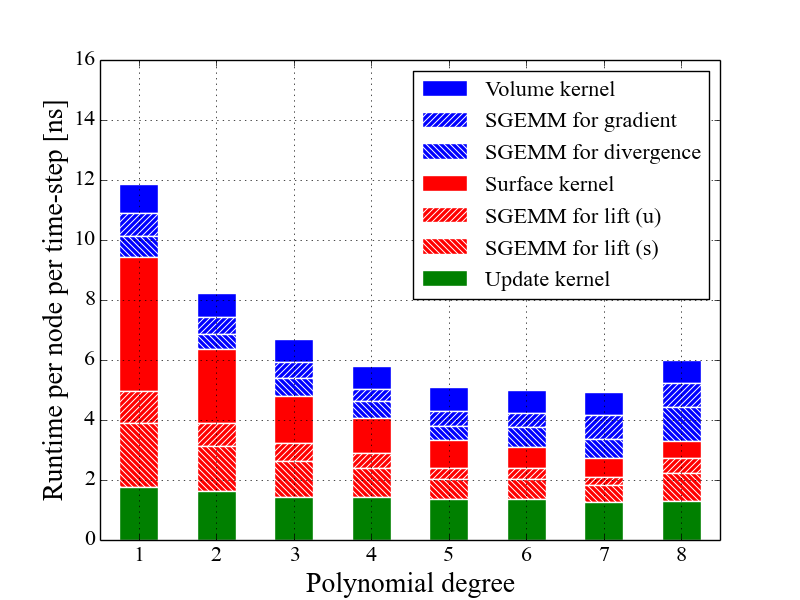}
    \label{fig:perf:runtimeElast:Cublas}
  \end{subfigure}
\end{tabular}
\caption{Global runtimes per node per time step with the three DG implementations in the elastic case (a), and runtimes per node kernel-by-kernel for each implementation (b)-(c)-(d).
\modif{For each polynomial degree, a mesh corresponding to approx. 2 millions of nodes has been used.}}
\label{fig:perf:runtimeElast}
\end{figure}

With the one-node-per-thread strategy, the surface kernel is the most expensive kernel for small polynomial degrees, while the volume kernel takes most of the runtime for large degrees (see figure \ref{fig:perf:runtimeAcou:NPT}).
For small degrees, the performance of the surface kernel is less efficient than for the volume kernel (see figure \ref{fig:perfNPT:arch}), and it requires more floating-point operations (not shown for the sake of brevity).
As $N$ increases, the percentage of achievable performance of both kernels decreases, and reaches a plateau around $23\%$ (volume kernel) and $30\%$ (surface kernel) in the throughput bounded region.
Over $N=5$, the volume kernel requires more floating-point operations than the surface kernel, which explains the larger runtime.
The one-node-per-thread update kernel, which is also used for the strategy with SGEMM, exhibits a constant and high percentage of achievable performance (around $70\%$).
The volume kernel of that strategy also has a similar performance (around $67\%$).
For the surface kernel, the performance is bad for small degrees, but it increases until a plateau around $52\%$ for $N\geq6$.

The performance results then are consistent with the observed runtimes: both volume and surface kernels perform better for small degrees with the one-node-per-thread strategy, while both surface kernel and SGEMM routine perform better for large degree for the strategy with SGEMM.
The profiled statistics of the kernels are quite good.
There is no register spilling, and both the number of registers per thread and the shared memory per thread block are reasonable.
Increasing the number of elements per thread block helps to improve the occupancy for small polynomial degrees.
The occupancy and the hit rate for global loads cannot explain the variations of performance that are observed.
We have obtained from NVIDIA's profiler \verb!nvprof! that most of the stalls occur because of data requests (metric \verb!stall_data_request!) and execution dependencies (metric \verb!stall_exec_dependency!), which are difficult to avoid.

The observations made for the acoustic case also apply to the elastic case, which requires more floating-point operations and memory transfers.
We show the global runtime per node for the three elastic implementations and the detailed kernel-by-kernel runtimes on figure \ref{fig:perf:runtimeElast}.
The simulations have been done on the same meshes.
The one-node-per-thread strategy remains the most efficient for small degrees.
The strategy with SGEMM is as efficient as the one-node-per-thread strategy at $N=7$, and is better for $N=8$.
The one-element-per-thread strategy is again very bad, with results far worse than before.
Since the operations require a larger local memory, register spilling occurs for $N$ smaller than with the acoustic case, which worse the results.

The one-node-per-thread strategy is only $1.8-2.25$ times slower than for the acoustic case, while the number of fields increases by a factor $2.25$.
This is party explained by the operational intensity that is larger in the elastic case for both volume and surface kernels.
The strategy with SGEMM is $2.14-2.49$ times slower than in the acoustic case.

\begin{figure}[!t]
\centering
\begin{tabular}{cc}
  \begin{subfigure}[b]{8cm}
    \centering
    \caption{Net arithmetic throughput \it (acoustic case)} \vspace{-0.5cm}
    \includegraphics[width=7.5cm]{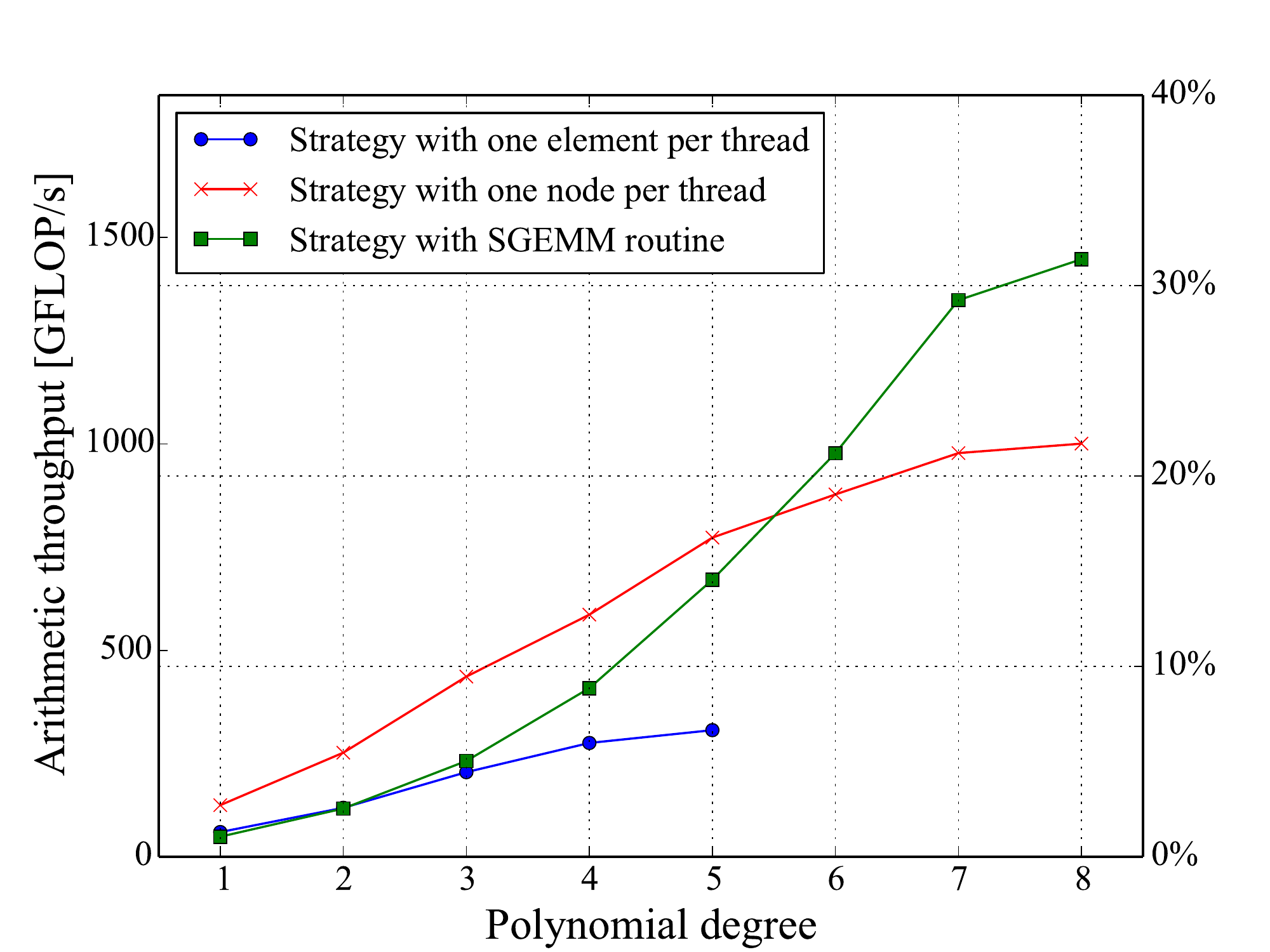}
    \label{fig:perfEPT:GFLOPs}
  \end{subfigure} &
  \begin{subfigure}[b]{8cm}
    \centering
    \caption{Net arithmetic throughput \it (elastic case)} \vspace{-0.5cm}
    \includegraphics[width=7.5cm]{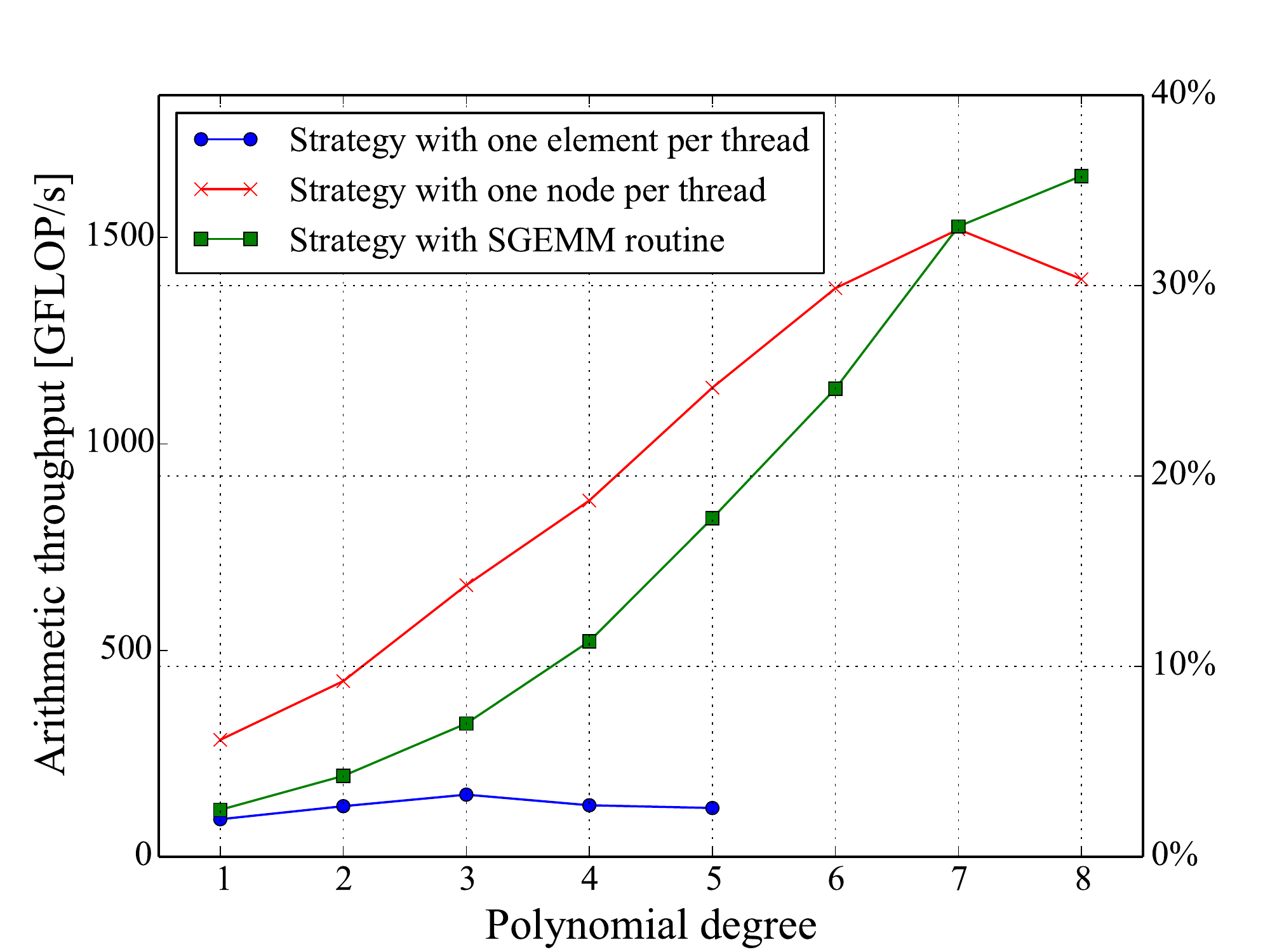}
    \label{fig:perfEPT:GFLOPsElast}
  \end{subfigure}
\end{tabular}
\caption{Net arithmetic throughput of the acoustic (a)-(b) and elastic (c)-(d) implementations.
  The net arithmetic throughput is obtained by dividing the total number of FLOP required per time step by the runtime required for one time step update.}
\label{fig:perf:netThroughput}
\end{figure}

Finally, to have an idea of the global performance of each strategy as a whole, we plot the net arithmetic throughput of each implementation for both acoustic and elastic cases in figure \ref{fig:perf:netThroughput}.
The net arithmetic throughput is obtained by dividing the total number of floating-point operations required for one full time step by the corresponding runtime required.
The implementation with SGEMM reaches 31.4\% of the peak throughput for the acoustic case, and 35.7\% for the elastic case.
A larger increase is observed with the one-node-per-thread strategy, from 21.7\% to 33\%.
This made of this strategy an interesting candidate for DG schemes requiring denser computations.




%% file: conclusion.tex
We have presented and analyzed three GPU implementations for time-domain discontinuous Galerkin simulations of acoustic and elastic waves.
Two of them are based on tailored kernels programmed by associating each thread to one element or one node, following strategies described by Kl\"ockner \etal \cite{Klockner2009} and Fuhry \etal \cite{Fuhry2014}, respectively.
We have also considered an alternative implementation with a specific partition of work which makes use of the SGEMM routine  of an external BLAS library.
All the implementations have been optimized and compared in the same computational framework with a three-dimensional benchmark.

The computational results highlight the strong performance of the one-node-per-thread kernels until $N=5$ for acoustic and $N=7$ for elastic waves.
The strategy with SGEMM performs better for higher polynomial degrees, though it requires more data transfers and more data storage in global memory.
The one-element-per-thread implementation does not work well for three-dimensional problems in our framework.
We have analysed both the SGEMM routine and the acoustic DG kernels using the roofline model, and we have identified bottlenecks for each computational strategy.

In the perspective of large-scale applications, it is essential to combine the selected implementation with a multi-rate time stepping scheme and parallel computation with several GPUs.
The one-node-per-thread implementation has already been adapted to multi-rate time stepping \cite{Gandham2015,Godel2010} and multi-GPU computation \cite{Modave2015}.
The implementation with SGEMM can be adapted in very similar ways.
In the future, we plan to investigate kernels for anisotropic wave propagation and artificial boundary conditions.